# Vortex Chains in Anisotropic Superconductors

Simon J. Bending[1] and Matthew J. W. Dodgson[2,3]

[1]Department of Physics, University of Bath, Claverton Down, Bath BA2 7AY, U.K.

[2]Institut de Physique, Université de Neuchâtel, Rue A. L. Breget, 2000 Neuchâtel, Switzerland

[3]Department of Physics and Astronomy, University College London, Gower Street, London WC1E 6BT, U.K.

## Abstract

High-$T_c$ superconductors in small magnetic fields directed away from the crystal symmetry axes have been found to exhibit inhomogeneous chains of flux lines (vortices), in contrast to the usual regular triangular flux-line lattice. We review the experimental observations of these chains, and summarize the theoretical background that explains their appearance. We treat separately two classes of chains: those that appear in superconductors with moderate anisotropy due to an attractive part of the interaction between tilted flux lines, and those with high anisotropy where the tilted magnetic flux is created by two independent and perpendicular crossing lattices. In the second case it is the indirect attraction between a flux line along the layers (Josephson vortex) and a flux line perpendicular to the layers (pancake vortex stack) that leads to the formation of chains of the pancake vortex stacks. This complex system contains a rich variety of phenomena, with several different equilibrium phases, and an extraordinary dynamic interplay between the two sets of crossing vortices. We compare the theoretical predictions of these phenomena with the experimental observations made to date. We also contrast the different techniques used to make these observations. While it is clear that this system forms a wonderful playground for probing the formation of structures with competing interactions, we conclude that there are important practical implications of the vortex chains that appear in highly anisotropic superconductors.

# 1  Overview

The vortex lattice that exists in a type II superconductor in a magnetic field is a fascinating example of the spontaneous appearance of a regular pattern not directly related to the underlying atomic structure. In this paper we will review the yet more complex structures that naturally occur when the magnetic field is tilted with respect to the symmetry axes of an anisotropic superconductor. We focus on the role that strong crystalline anisotropy plays in controlling the structure of vortex matter in layered superconductors like the high-$T_c$ cuprates. In particular the phenomenon of vortex chain formation is examined in two distinct limits; the cases of moderate and high anisotropy. As will be demonstrated the underlying physical mechanisms are very different in the two cases; being due to an attractive component of the interaction between tilted flux lines in the first case, and the formation of two independent and perpendicular crossing lattices in the second. The observed physical phenomena are found to be especially rich in highly anisotropic materials.

This paper is organised in the following manner. Section 2 presents a brief introduction to vortex matter and introduces the role played by varying degrees of crystalline anisotropy. Section 3 focuses on superconductors with moderate anisotropy and presents the theoretical understanding of chain formation in these materials along with descriptions of experimental observations of them. Section 4 treats highly anisotropic superconductors. It begins with a description of pancake and Josephson vortices in layered materials, and then extends these concepts to achieve a theoretical understanding of interacting crossing lattices and pancake-vortex chain formation under tilted magnetic fields. The remainder of this section describes experiments which have revealed the existence of crossing lattices and vortex chains in the high anisotropy limit, as well as studies of the rather remarkable dynamic properties of this state of interacting vortex matter. Finally section 5 contains some concluding remarks and a brief discussion of potential future developments in this fascinating scientific area.

## 2 A Brief Introduction to Vortex Matter

Understanding the behaviour of most known superconductors requires a thorough study of the physics of quantised vortices. The "mixed-state" of type-II superconductors allows for a partial penetration of magnetic fields by introducing vortex lines, each of which carries a flux quantum of $\Phi_0 = hc/2e$. The magnetic response of this state is therefore determined by the vortex interactions: their mutual repulsion is balanced by the "pressure" of the external magnetic field.

In conventional, isotropic superconductors, there are two important length scales: the London penetration depth, $\lambda$, for the variation of currents and magnetic fields, and the coherence length, $\xi$, which sets the scale for variations in the amplitude of the macroscopic wave function, or equivalently, the order parameter of the superconductor. These are the two length scales that enter the Ginzburg-Landau equations which successfully describe the phenomenology of superconductors (see e.g., [1,2]). A type-II superconductor is defined by $\lambda/\xi > 1/\sqrt{2}$, and Abrikosov discovered that, in this case, the mixed state is a solution of the Ginzburg-Landau equations [3]. This theoretical prediction of vortices recently earned Abrikosov the Nobel Prize along with Ginzburg. These vortices contain a core on the scale of $\xi$ where the order parameter amplitude falls to zero. Around this core the phase of the order parameter winds by $2\pi$, and currents flow proportional to the gradient in phase as far as $\lambda$ from the vortex centre. Beyond $\lambda$ the magnetic field and currents decay exponentially.

The mixed state is of crucial technological importance in that it greatly extends the range of magnetic fields where the superconducting state can exist: In a type-I superconductor the upper critical field is proportional to $\Phi_0/\lambda \xi$, while the mixed state has an upper critical field greater by a factor of $\sqrt{2}\lambda/\xi$. As well as determining the magnetization, the vortices affect the current-voltage response: The vortex lines feel a force due to passing currents, but their motion will cause dissipation. Therefore a superconducting response is only found when the vortex cores are effectively pinned by inhomogeneities of the material.

The phenomenon we will look at in this review is the formation of chains of vortex lines in anisotropic type-II superconductors when the magnetic field does not lie along a symmetry axis. The main family of superconductors where these chains have been observed are the high-$T_c$ cuprates. In these layered materials, the supercurrents mainly flow within the 2D (two-dimensional) $CuO_2$ layers, with a weaker tunnelling current between the layers. The geometry is defined in Fig. 2.1. It is found that the penetration depth for currents perpendicular to the layers, $\lambda_z$, is much greater than that for currents within the layers, $\lambda_\parallel$. We will characterize this anisotropy by the ratio $\gamma = \lambda_z/\lambda_\parallel$ and typical values range from $\gamma \approx 8$ for $Y_1Ba_2Cu_3O_{7-\delta}$ (YBCO) up to as much as $\gamma \approx 500$ for $Bi_2Sr_2CaCu_2O_{8+\delta}$. (BSCCO). For some of the cuprates, (e.g. YBCO) the existence of CuO chains along the a axis means that there is also a slight difference between $\lambda_a$ and $\lambda_b$.

The layered structure leads to some important differences between a flux line directed perpendicular or parallel to the layers. In the first case, the phase singularity of the vortex, and therefore its normal core, only exists in the superconducting layers. One can therefore think of the flux line as a stack of 2D vortices, or "pancake vortices". However, the flux line along the layers hides its phase winding around the normal region between two neighbouring layers. This is reminiscent of the situation in a 2D Josephson junction, and therefore we call this in-plane flux line a Josephson vortex. Pancake vortices and Josephson vortices are described in more detail in section 4.1. More thorough descriptions are available in several reviews that have been written on the theory of vortices in high-$T_c$ superconductors [4-8].

The remainder of this article treats separately two classes of superconductor: "moderately" and "highly" anisotropic. The criterion for these two labels is set by whether $\gamma$ is smaller or larger than $\lambda_\parallel / d$. This is important because the nature of the coupling between pancakes in different layers is different in each class: for moderately anisotropic superconductors the Josephson energy of phase differences from layer to layer dominates the coupling, while in the high anisotropy limit the long ranged magnetic coupling between pancakes is most important. For our purposes this is apparent in that a different type of chain has been observed in each class. For the moderately anisotropic superconductors, the layeredness is not so important, and the vortices can largely be described in terms of the theory of continuous anisotropic superconductors. This is described in section 3. On the other hand, the highly anisotropic superconductors allow for a new type of vortex structure where the tilted magnetic field is made from almost independent contributions from JV's and PV stacks. These crossing lattices, and the resulting chains of PV stacks, are described in section 4. A schematic diagram of the geometry considered in this paper is shown in Fig. 2.1.

## 3 Tilted Chains in Moderately Anisotropic Superconductors

## 3.1 Theory of Vortex Chains in Moderately Anisotropic Superconductors

The starting point for the description of vortices in type-II superconductors is the phenomenological Ginzburg-Landau theory. In this theory, the free energy of the superconductor is expanded in terms of a complex order parameter, $\psi(\mathbf{r}) = |\psi(\mathbf{r})| e^{i\phi(\mathbf{r})}$, and its derivatives. Minimizing this free energy with respect to the order parameter and the vector potential, $\mathbf{A}(\mathbf{r})$, leads to the well-known Ginzburg-Landau equations. Abrikosov solved these equations for type-II superconductors to find the mixed state where there is penetration by vortex lines.

As long as the vortices are separated by larger distances than the vortex core size, we can ignore variations in the amplitude of the order parameter, $|\psi(\mathbf{r})|$, and write the supercurrent in terms of the vector potential and the phase,

$$\mathbf{j} = -\frac{c}{4\pi\lambda^2}\left(\mathbf{A} + \frac{\Phi_0}{2\pi}\nabla\phi\right). \qquad (3.1.1)$$

This is known as the London model [taking the curl of (3.1.1) leads directly to the London equation. Note that the current is gauge-invariant: a gauge transformation in A must be accompanied by the corresponding transformation in the phase $\phi$ to keep $\mathbf{j}$ fixed.]. The generalization to an anisotropic superconductor is straightforward: a different value of the penetration depth $\lambda_i$ must be used along the different symmetry axes,

$$j_i = -\frac{c}{4\pi\lambda_i^2}\left(A_i + \frac{\Phi_0}{2\pi}\nabla_i\phi\right). \qquad (3.1.2)$$

The anisotropic London model has a free energy depending on the field and currents of the form,

$$\begin{aligned} F_L &= \frac{1}{8\pi}\int \mathrm{d}^3 r \left[ B^2 + \sum_i \frac{(4\pi\lambda_i)^2}{c^2} j_i^2 \right] \\ &= \frac{1}{8\pi}\int \mathrm{d}^3 r \left[ |\nabla\times\mathbf{A}|^2 + \sum_i \frac{1}{\lambda_i^2}\left| A_i + \frac{\Phi_0}{2\pi}\nabla\phi \right|^2 \right]. \end{aligned} \qquad (3.1.3)$$

At this stage we note that the anisotropic London relation (3.1.2) can be cast in the isotropic form of (3.1.1) by the rescaling procedure [9,10],

$$\tilde{r}_i = \gamma_i r_i, \text{ and } \tilde{A}_i = A_i / \gamma_i. \qquad (3.1.4)$$

with $\gamma_i = \lambda_i / \lambda$ (where the choice of $\lambda$ is arbitrary). This also makes the second term in the London energy isotropic. This anisotropic rescaling procedure is therefore valid either when we can ignore variations in the magnetic field (dropping the gradient in **A** versus **A**) or when the field is directed along a symmetry axis. In these cases, there is no new physics associated with anisotropic superconductors: all the relevant features are recovered from the isotropic case after rescaling.

An immediate consequence of the rescaling theory is that the flux-line lattice in an anisotropic superconductor should be distorted: If the field is directed along a symmetry axis, then the rescaling theory applies even to an isolated flux line, which now has the structure of a conventional flux line, but with the field and current having elliptical profiles, decaying over the length scales $\lambda_i$ and $\lambda_j$ along the two remaining symmetry axes, $i$ and $j$. Then, at any finite mean magnetic field, the resulting vortices should form a stretched triangular lattice with a ratio of $\gamma_i / \gamma_j$ between the lattice constants along the respective axes. At field angles not along a symmetry axis, the same effective stretching of the regular triangular lattice occurs as long as the flux density is so high that the fields of the flux lines are strongly overlapped.

New physics may occur, however, for moderate magnetic fields where the flux lines do not overlap too closely, as long as the field is at a finite angle from the symmetry axes. This is where the vortex chains occur that form the focus of this article. We now consider the structure of the vortex system at tilted angles, assuming that all flux lines are aligned in the same direction. In principle the linearity of the London model simplifies things greatly: we only need to solve the structure of one vortex line, and a finite density of vortices corresponds to a superposition of these solutions. The energy of interaction between the flux lines is simply given by the value of the field from one flux line at the centre of a second flux line. However, the single flux line has a very complex general solution. This was first noted by Kogan [11] who showed that as the currents are no longer restricted to lie in the plane perpendicular to the vortex line, there must be field components perpendicular to the vortex direction. These components average to zero when integrated over all of the volume of the flux line. Even more surprising was the result obtained in later studies for the case of uniaxial anisotropy [12,13] which showed that the component of magnetic field in the vortex direction changes sign at large distances from the vortex centre along the *x* direction (if the vortex is tilted within the *x-z* plane; in the *y* direction this component has no sign change). This also implies that that two tilted flux lines may have an attractive interaction at large distances! The physical consequence, emphasized by Buzdin and Simonov, is that a tilted magnetic flux will first penetrate as chains of finite density, with the interaction between the flux lines optimised between the short-range repulsion and the long-range attraction.

In Figure 3.1.1 we demonstrate the changeover from the rescaled lattice to the chain state in a layered superconductor with moderate anisotropy. This figure shows the results of a numerical minimization of the interaction energy of a tilted flux-line lattice with respect to its lattice parameters for different values of the out of plane field, $B_z$, and the angle, $\theta$, of the field from the *z* axis [8]. Plotted is the lattice spacing in the tilt direction, $R_1^x$. The second lattice vector in the *x-y* plane is then given by, $\vec{R}_2 = \left( \dfrac{R_1^x}{2}, \dfrac{\Phi_0}{B_z R_1^x} \right)$. For $\theta = 0$ we have $R_1 = R_2 = \left[ (2/\sqrt{3})\Phi_0 / B_z \right]^{1/2} \equiv a_\Delta$.

We see that for high values of magnetic field the lattice is given by rescaling theory, which always has $R_1^x$ larger than $R_2$. However, at low fields the opposite distortion is found, due to the formation of chains with a small value of $R_1^x$. In this limit, the value of $R_1^x$ is even smaller than the optimum separation in an isolated chain, due to the repulsive interaction between chains (a larger chain separation demands a smaller vortex spacing within the chain if the density is fixed).

**3.2  Experimental Observations of Vortex Chains in Moderately Anisotropic Superconductors**

The first experimental observation of an ordered Abrikosov flux-line lattice in a high-temperature superconductor was made by the Bitter decoration method [14]. This technique involves cooling the sample in the desired applied field to low temperature (often 4.2K) and then evaporating over it a 'smoke' of ferromagnetic particles of typical diameter 5-10nm. These particles will follow the maximum field gradient down towards the sample surface, and hence will migrate towards the cores of any vortices present. Once there the particles adhere to the surface via weak van der Waals forces, and the sample can be warmed up to room temperature and viewed in a scanning electron microscope without losing the magnetic information. Gammel *et al*. [14] were able to show that the vortices in a YBCO single crystal ($T_c$=92K) formed a well-ordered triangular lattice which was indistinguishable from that observed earlier in conventional superconductors. YBCO has a small orthorhombic distortion of the crystalline lattice (a≠b), and as a consequence the vortex lattice should show a weak anisotropy in the a-b plane, as we explained in section 3.1. This was not observed in the first Bitter decoration experiments as the crystals used were strongly twinned (the coexisting domains of different ab orientations lead to isotropic superconductivity on length scales larger than the domain size). Subsequent Bitter decoration experiments on single crystals with large twin-free regions [15-17] or epitaxial thin films [18] later revealed this anisotropy, which is quite small in YBCO where $\lambda_b/\lambda_a$=1.15±0.02. Such a small effect is difficult to pick up directly from Bitter decoration images or from their Fourier transforms. The anisotropy of $Y_1Ba_2Cu_4O_8$ (YBCO-124) is considerably larger and Fig. 3.2.1 shows a Bitter image of a high quality YBCO-124 single crystal ($T_c$~80K) (crystal grown by Prof. J.Karpinski, ETH, Zürich; Bitter decoration by Prof. L.Ya.Vinnikov, Institute of Solid State Physics RAS, Chernogolovka, Russia) after field cooling to low temperature in $H_z$=41Oe. A digital Fourier transform of the pattern is inset in the top corner of the figure, which clearly reveals a strong distortion to the quasi-triangular lattice and an anisotropy of about 1.4±0.05 in the a-b plane. Dolan *et al*. [15] were even able to successfully generate Bitter decoration images in the highly anisotropic a-c plane in YBCO. Fig. 3.2.2 shows such an image of an YBCO single crystal after field cooling to low temperatures in $H_\parallel$=4Oe. The image shows the presence of oval vortices which are highly elongated in the in-plane direction (perpendicular to the c-axis). These oval vortices in turn form a highly anisotropic lattice; dense chains of vortices form along the c-axis (mean intrachain spacing ~1.2μm, mean interchain spacing ~6.7μm), with an estimated anisotropy of , $\lambda_a/\lambda_c \cong \lambda_b/\lambda_c$~5.5±1.0. This image is a dramatic demonstration of the rescaling theory (3.1.4), which should be valid in this case as the field is along a symmetry axis.

Bitter decoration was also used to make the first observation of the chain state in YBCO single crystals under tilted magnetic fields [19]. Fig. 3.2.3 shows an image of the surface of a YBCO crystal which has been decorated after field-cooling to low temperature in a total field of 24.8Oe applied at 70$^{\circ}$ away from the c-axis. The formation of high-density vortex chains along the direction of the in-plane field (in this case approximately 45$^{\circ}$ away from the a-axis) is clearly demonstrated. Experimentally the intrachain separation was only weakly dependent on the perpendicular component of magnetic field, $H_z$. This was in reasonable agreement with earlier theories of anisotropic vortex lattices based on the London equation [13] which predicts an attractive well in the vortex-vortex interaction along the chains. In the limit $H \sim H_{c1}$ the vortices just sit in these wells regardless of the magnitude of $H_z$, yielding an interchain spacing $\sim 1/B$ in order to conserve net flux density. At higher fields inter-chain interactions come into play (as shown in section 3.1) and the intrachain spacing is no longer constant. In order to avoid this issue Gammel *et al*. studied the intrachain spacing as a function of tilt angle at fixed $H_z$=12Oe, and found a non-monotonic dependence in reasonable quantitative agreement with theory. Observation of chain formation was successful in the tilt angle range between 40$^{\circ}$-70$^{\circ}$. At low applied fields ($H_z \sim$4Oe) quenched disorder led to an amorphous vortex array, while at high fields ($H_z$>20Oe) the chains merged slowly and smoothly to form an ordered triangular lattice. Indeed the defects (predominantly edge dislocation) which are clearly visible in Fig. 3.2.4 appear to be the first step in this merging process.

In the theory of Buzdin *et al*. [13] the vortex chains are composed of vortices which are uniformly tilted with respect to the crystalline c-axis throughout the sample. This has been verified in Lorentz microscopy images of very thin (300-400nm) YBCO single crystals [20]. In these experiments the tilted sample was illuminated with a beam of 1MeV electrons from a bright field emission source. Electrons passing through the high field region near the core of a vortex are deflected by the Lorentz force there in such a way that the electron flux is enhanced near one edge of the fluxon and depleted near the other. Hence a vortex can be recognised as an oval region with adjacent halves of bright and dark contrast. Fig. 3.2.4 shows a series of Lorentz micrographs of a YBCO crystal at $T$=30K and fixed $B_z$=3Oe, at various field tilt angles between 75$^{\circ}$ and 84$^{\circ}$. As the tilt angle is increased the vortex images begin to elongate along the direction of $H_{\parallel}$ and above 80$^{\circ}$ the formation of linear chains is observed. This is conclusive evidence that these chains of vortices in YBCO are formed by an attractive interaction between uniformly tilted vortex lines. As we will see in the next section this is in stark contrast to the situation in much more anisotropic BSCCO crystals.

## 4 Chains due to crossing lattices in highly anisotropic superconductors

### 4.1 Vortices in layered superconductors

There are situations where the discrete coupling in a layered superconductor has an important effect on the physics of vortices induced by magnetic fields (see [8] for a detailed review). In particular, this occurs when a significant phase difference can build up between local points in neighbouring layers. This is usually treated phenomenologically with the Lawrence-Doniach model [21,22], which is basically an extension of the usual Ginzburg-Landau model, but with a Josephson-like term representing the energy of phase differences from layer to layer. In the London limit of fixed amplitudes of the superconducting order parameter, the continuous model (3.1.2) becomes replaced by a layered version, which results in the following relations between current and phase,

$$\vec{j}_{\parallel}^{n} = -\frac{c}{4\pi\lambda_{\parallel}^{2}}\left(\vec{A}_{\parallel}^{n} + \frac{\Phi_{0}}{2\pi}\vec{\nabla}_{\parallel}\phi_{n}\right), \qquad (4.1.1a)$$

$$j_{z}^{n,n+1} = -\frac{c\Phi_{0}}{8\pi^{2}\lambda_{z}^{2}d}\sin\varphi_{n,n+1}. \qquad (4.1.1b)$$

Here, $\phi_{n}$ is the phase in the $n$th layer and $\varphi_{n,n+1}$ is the gauge-invariant phase difference between neighbouring layers. The layer separation enters as $d$. The in-plane currents therefore have the same relation as for a continuous superconductor (3.1.2), while the $z$-axis current obeys the non-linear Josephson relation. Notice that in the limit of large anisotropy, $\lambda_{z} = \gamma\lambda_{\parallel} \rightarrow \infty$, the z axis current vanishes.

For a flux line directed along the $z$ axis, the currents all flow within the layers. They therefore all obey (4.1.1a), and there is little difference in the structure from that of a flux line in a continuous superconductor. However, the vortex core (where the phase singularity forces a suppression of the superconducting order parameter) now strictly only exists in the layers, rather than along a continuous line, and so one can think of the flux line as made from a stack of two-dimensional vortices, see figure 4.1.1(a). Due to the wide circular shape of their current distribution, these were labelled "pancake vortices" (PVs) by Clem [23]. Distortions of the flux line can now be described in terms of displacements of the PVs, and the properties of a PV have been described in several theoretical works [24-26]. At finite magnetic fields along the $z$ axis, a finite density of PV stacks is induced. Assuming there is no anisotropy for currents within the layers, these stacks then form a regular triangular lattice.

The flux lines induced at fields along the layers are different from PV stacks for two reasons: the anisotropy of the circulating currents which must flow both within and perpendicular to the layers, and the discreteness of the layers which allows the phase "singularity" to lie in the non-superconducting region between two neighbouring layers. There is then no need for the drastic suppression of the order parameter that exists at the centre of a PV, as long as the superconducting coherence length in the $z$ direction, $\xi_z = \xi_{\parallel}/\gamma$, is much smaller than the layer spacing, $d$, which is the case for all of the high-$T_c$ cuprates. As the phase winding at the centre of this flux line is quite similar to that in a two-dimensional Josephson junction, the in-plane flux lines are known as Josephson vortices (JV's). At the "core" of the JV, the phase difference across the central junction winds from 0 to $2\pi$ over the length scale $\gamma d$, which is known as the Josephson length. The currents around this core are schematically shown in figure 4.1.1(b). Away from this core, the phase differences are small, and the non-linear relation (4.1.1b) can be replaced by the linear limit (3.1.2), with $\lambda_i = \lambda_z$. Therefore the current and magnetic fields in the JV resemble at large distances those of a flux line in a continuous anisotropic superconductor, as described above in section 3.1. For this reason, for not too large in-plane magnetic fields $B_x \ll \Phi_0/\gamma d^2$, the JV lattice is just a stretched version of the usual triangular lattice.

As we described in section 3.1, the vortex lattice directed along a symmetry axis in the continuous limit is just stretched according to the scaling theory (3.1.4), see e.g., [27]. However, there is still the rotational degeneracy of the triangular lattice in the scaled coordinates, which means there are many possible orientations of the Josephson vortex lattice. For the purpose of this review we will only consider the orientation which is formed from "stacks" of JV's perpendicular to the layers, and with the smallest JV separation within each stack. The reason for this choice will become apparent in section 4.2. For now we note that this orientation has a lattice constant in the $z$ direction of size,

$$a_z = \sqrt{\frac{2\Phi_0}{\sqrt{3}\gamma B_x}}\,, \tag{4.1.2}$$

and a separation between the JV stacks of,

$$a_y = \sqrt{\frac{\sqrt{3}\gamma\Phi_0}{2 B_x}}\,. \tag{4.1.3}$$

Of course there are only a discrete set of fields, $B_x$, for which $a_z$ is an integer multiple of the layer spacing. Therefore, the lattice at a typical field will have some incommensuration effects, and the above values of $a_z$ and $a_y$ should only be taken as an estimate. For an in-plane field of $B_x = 50\,\text{G}$, a layer spacing of $d = 15\,\text{Å}$, and an anisotropy ratio of $\gamma = 500$ we will have $a_z \approx 20d$, and so incommensuration may lead to a variation of $\sim 3\%$ in the lattice parameter.

We conclude this subsection with a review of the structure of the Josephson vortex "core". In the central junction of the vortex, and at distances less than $\gamma d$ from its centre, the phase difference is large enough that the non-linearity of (4.1.1b) must be included. To find the actual phase configuration, one must apply the current conservation law, $\nabla_\| \cdot \mathbf{j}_\| + 1/d \left( j_{n,n+1}^z - j_{n-1,n}^z \right) = 0$, to (4.1.1) to find a single non-linear differential equation for the gauge invariant phase difference $\varphi_{n,n+1}$. Now, as long as we are not too far from the vortex centre (more specifically, when $y^2 + \gamma^2 z^2 < \lambda_z^{\,2}$) we can ignore the vector potential contribution to $\varphi_{n,n+1}$, and then the phase of the Josephson vortex is a solution to,

$$\frac{d^2\phi_n}{dy^2} + \frac{1}{(\gamma d)^2}\Big[ \sin\left( \phi_{n+1} - \phi_n \right) - \sin\left( \phi_n - \phi_{n-1} \right) \Big] = 0 . \quad (4.1.4)$$

This must be solved numerically subject to the phase-winding boundary conditions of the JV, that $\phi_n$ winds from $-\pi$ to zero for increasing $y$ at $n \geq 1$, and from $+\pi$ to zero for increasing $y$ at $n \leq 0$. A variational solution valid also at large distances where the currents become screened was found by Clem and Coffey [28]. This was generalized by Koshelev, who also found the numerical solution using a relaxation method [29]. For our purposes we note here that the in-plane current which, in the validity range of (4.1.4) is just $j_{\|,n} = -\dfrac{c\Phi_0}{8\pi^2 \lambda_\|^{\,2}} \dfrac{d\phi_n}{dy}$, has the following behaviour immediately above the centre of the JV,

$$j_{\|,n} \approx -\frac{c\Phi_0}{8\pi^2 \lambda_\|^{\,2} \gamma d} \frac{C_n}{\left( n - \frac{1}{2} \right)}, \qquad (4.1.5)$$

where $C_1 \cong 0.55$, $C_2 \cong 0.86$, $C_3 \cong 0.95$ and $C_n \cong 1$ for large $n$, where the linear approximation is valid. Note that for $|n| > \lambda_\|/d$, screening is important and the current will fall off exponentially with $n$.

## 4.2 Theory of Interacting Crossing Lattices

It is now well established that, for layered superconductors with large anisotropy, there is an alternative arrangement of flux lines in the presence of a tilted magnetic field: a coexistence of two perpendicular lattices that cross through each other. Speculations of such a configuration go back over a decade [30,31], and indeed, a connection between the observation of chains in BSCCO [32] and the crossing-lattice state was made by Huse [33]. A more quantitative picture emerged after the work of Koshelev [34], where the distortions from simple straight crossing lattices were first calculated.

Koshelev considered the forces on a pancake vortex within a PV stack when the stack crosses through the centre of a Josephson vortex. The in-plane currents of the JV exert a Lorentz force on the PV in a direction perpendicular to the current with magnitude $f_{\text{Lorentz}} = \dfrac{\Phi_0 d}{c} j_{\parallel}$ (see e.g. [8]). This will tend to displace the PV away from the stack until the force is balanced by the interaction of the PV with the rest of the stack. In the limit of very large anisotropy, we can write the PV interactions in a pair-wise form given by the magnetic coupling. The equilibrium distortion is then given when the forces are balanced, which is determined by the self-consistent relation,

$$\frac{\Phi_0 d}{c} j_{\parallel,n} = \sum_{m \neq n} f_{\text{p}}^{m-n}\left(u_m - u_n\right), \qquad (4.2.1)$$

where $f_{\text{p}}^{m-n}\left(u_m - u_n\right)$ is the magnetic force on the $n$th pancake from the $m$th pancake. The in-plane current of the JV, $j_{\parallel,n}$, is given by the above formula (4.1.5). This rather complex self-consistent problem was simplified by Koshelev by invoking the long range of the magnetic interaction between pancakes in different layers to ignore the effect of the displacements of other pancakes on the total force experienced. In addition he expanded for small displacements to find an *almost* quadratic "cage" potential for each pancake in the stack. The result is a restoring force for each pancake of the form,

$$\sum_{m \neq n} f_{\text{p}}^{m-n}\left(u_m - u_n\right) \approx u_n \frac{\Phi_0^2 d}{(4\pi)^2 \lambda_{\parallel}^4} \ln\left(\lambda_{\parallel}/r_{\text{w}}\right), \qquad (4.2.2)$$

where $r_{\text{w}}$ is a measure of the maximum PV displacement in the JV core, to be determined. Equating this force to the Lorentz force as in (4.2.1) gives the result for the lateral displacement of the $n$th PV above a given JV core, [34],

$$u_n \approx \frac{2 C_n \lambda_{\parallel}^2}{(n-1/2)\gamma d \ln\left(\lambda_{\parallel}/r_{\text{w}}\right)}, \qquad (4.2.3)$$

where $C_n$ is given above in (4.1.5).  To logarithmic accuracy we can therefore write the displacement cut-off as $r_w = \lambda_\parallel^2 / \gamma d$ (this procedure only makes sense in the very large anisotropy limit when $\gamma d \gg \lambda_\parallel$ ).  The crossing configuration derived here is shown in figure 4.2.1.

The energy of the equilibrium crossing configuration, in comparison to that of a straight PV

stack, is found by integrating the force and summing over all of the pancakes, which in the cage approximation gives [35],

$$E_\times \approx -\frac{\Phi_0^2}{2\pi^2 \gamma^2 d \ln\left(3.5\gamma d / \lambda_\parallel\right)} \ . \qquad (4.2.4)$$

This negative energy represents an indirect attraction of PV stacks to the centre of a JV.  In fact, one can perform a similar calculation with the PV stack a distance $y$ from the JV stack centre, to find a profile of this interaction energy, see figure 4.2.2.  This is the mechanism for the formation of high density chains of PV stacks along the stacks of JVs that are widely separated in the Josephson vortex lattice.

It is important to consider the approximations used above.  The simple step of going beyond the quadratic cage potential was performed by one of us [36], by writing the full interaction of a pancake with the rest of a straight stack.  This has the interesting feature that at large distances,

$u_n \gg \lambda_\parallel$, the restoring force decays as $1/r_\parallel$ , which will eventually be smaller than the Lorentz force,

implying that the pancake is peeled off from the stack at large distances.  A consequence is that there is a critical anisotropy below which the crossing configuration is unstable.  This approach finds a bound on the stability limit of $\gamma > \gamma_c \approx 2.86\, \lambda_\parallel / d$ .  However, as the instability is reached when the largest displacement, $u_1$, approaches $\gamma d$ , it is important to consider also the relaxation of

Josephson currents.  This requires a full numerical solution for the phases due to the non-linearity of the Josephson term.  Recently such a calculation was performed by Koshelev [37], with the result for the stability limit of $\gamma_c \approx 1.45\, \lambda_\parallel / d$ , i.e. there is a larger range of parameters for which crossing remains stable.  It is of interest that the same numerical calculations show the cage approximation to give good results for the displacements and energy as long as $\gamma > 2.9\, \lambda_\parallel / d$ .  This justifies the neglect of Josephson coupling for large enough anisotropy.  For smaller anisotropy,

$\gamma_c < \gamma < 2.9\, \lambda_\parallel / d$ , the energy becomes more negative than the result in (4.2.4) by as much as 30%.

Having established the stability of a Josephson vortex crossing a PV stack, we may ask whether or not the crossing lattices provide a preferable configuration to a simple tilted flux line lattice.  A simple argument for a transition between these two possible states as a function of field angle was provided by Koshelev [34]. For small tilt angles away from the $z$ axis, the energy of the tilted lattice is given in terms of the tilt modulus, $c_{44}$, as,

$$E_{tl}\left(B_x, B_z\right) = E\left(B_z\right) + \frac{1}{2}c_{44}\left(\frac{B_x}{B_z}\right)^2 . \qquad (4.2.5)$$

In contrast, the energy of the crossing lattice for a small tilt angle is linear in the in-plane field,

$$E_{cl}\left(B_x, B_z\right) = E\left(B_z\right) + \frac{B_x^2}{8\pi} + \frac{B_x}{\Phi_0}\varepsilon_{JV} , \qquad (4.2.6)$$

where $\varepsilon_{JV}$ is the energy per unit length of a Josephson vortex, which may be renormalized by the presence of the PV stacks. Clearly, the quadratic-versus-linear dependence on $B_x$ means that there should be a first-order transition from a tilted lattice to crossing lattices on increasing the in-plane field from zero. To get an order-of-magnitude estimate, we can take the magnetic coupling contribution to the tilt modulus when $B_z > \Phi_0/\lambda_\parallel^2$, which is [38],

$$c_{44} = \frac{B_z^2}{4\pi} + \frac{3.68\Phi_0^2}{\left(4\pi\lambda_\parallel\right)^4} , \qquad (4.2.7)$$

and the JV energy in the absence of PV stacks [28,29],

$$\varepsilon_{JV} = \frac{\varepsilon_0}{\gamma}\ln\left(4.66\lambda_\parallel/d\right) , \qquad (4.2.8)$$

to give a transition at,

$$B_x^{tl\to cl}\left(B_z\right) \approx \frac{B_z^2}{\gamma}\frac{\left(4\pi\lambda_\parallel\right)^2}{\Phi_0} . \qquad (4.2.9)$$

This leads to a small value of the tilt angle at the transition for large anisotropy and moderate values of $B_z$.

The assumptions of this simple derivation break down for large tilt angles, which will be the case when $B_z > \frac{\gamma\Phi_0}{\left(4\pi\lambda_\parallel\right)^2}$, and for large enough in-plane fields that the JVs start to interact, when $B_x > \frac{\Phi_0}{\gamma\lambda_\parallel^2}$, or equivalently, $B_z > \frac{\Phi_0}{\left(4\pi\lambda_\parallel\right)^2}$. In addition, there are further complications when the crossing lattices respond to their mutual interaction, leading to inhomogeneous structures (see below). A full numerical analysis of the phase boundary between crossing lattices and the tilted lattice has not yet been carried out, although earlier works [39,40] together with some more exact recent calculations [41] have shown quite complex behaviour: e.g., for not too high an anisotropy the phase boundary in the $B_x$-$B_z$ plane is non-monotonic, and there is a re-entrant transition back to the tilted lattice on further increasing $B_x$. This is not so unexpected, as it was already pointed out [29] that adding a small component of $B_z$ to the Josephson vortex lattice at finite $B_x$ may induce single PV kinks in the JV's rather than complete PV stacks. This scenario occurs when the kink energy is less than the PV stack energy per pancake, which is the case for not too large anisotropy.

As first postulated by Huse, and quantified by Koshelev, the attraction between PV stacks and JVs may lead to the formation of chains of PV stacks with density higher than the average of the PV lattice. In fact, such chains will only form if the attraction is sufficient to overcome the energy cost in bringing the repelling PV stacks nearer. For large values of $B_z$ there is a high density of PVs and the chains do not form, whereas at very low $B_z$ there is nothing to prevent the formation of chains. Koshelev proposed that there is a phase transition between an undistorted PV lattice and a composite state of PV lattice separating the PV chains of a higher density [34]. He estimated the transition point by comparing the energy to form an interstitial defect in the PV lattice to the energy gain from sitting on top of a JV stack. An alternative approach is to consider the incommensuration energy of the chain with slightly higher density than the surrounding lattice. In fact, both approaches give parametrically the same transition field in the dilute limit: The interaction energy is determined by the exponential tail of nearest neighbours, $\varepsilon_{int} \sim \varepsilon_0 e^{-a_\Delta / \lambda_\parallel}$, where $\varepsilon_0 = \left( \Phi_0 / 4\pi\lambda_\parallel \right)^2$ is the energy scale per unit length of the PV stack, and $a_\Delta = \sqrt{2\Phi_0 / \sqrt{3}B_z}$ is the spacing of the PV lattice. Comparing this to the attractive energy per unit length of the PV stack to the JV stack, $\varepsilon_{pin} = E_\times / a_z$, gives a transition field of,

$$B_z^{pure \rightarrow comp} \approx \frac{\Phi_0}{\lambda_\parallel^2} \Bigg/ \ln\left[ A\left( \frac{\gamma d}{\lambda_\parallel} \right)^2 \left( \frac{\Phi_0}{\gamma d^2 B_x} \right)^{\frac{1}{2}} \right]. \qquad (4.2.10)$$

Similarly, on decreasing $B_z$ further there is a transition from the composite lattice+chain state to an isolated-chain state, which is found by comparing the PV-stack-to-JV-stack attraction to the repulsion of the PV stacks along the chain. Now, the energy change of removing one PV stack from a chain with nearest neighbour spacing $a_{ch}$ is

$$\delta\varepsilon = -\varepsilon_{pin} - 4\varepsilon_0 \sqrt{\frac{\pi\lambda_\parallel}{2a_{ch}}} e^{-a_{ch}/\lambda_\parallel}, \qquad (4.2.11)$$

where we have used the exponential tail of the flux line interaction energy, see e.g., [2]. This energy change is zero when

$$a_{ch} \approx \lambda_\parallel \ln\left[ C\left( \frac{\gamma d}{\lambda_\parallel} \right)^2 \left( \frac{\Phi_0}{\gamma d^2 B_x} \right)^{\frac{1}{2}} \right], \qquad (4.2.12)$$

which corresponds to a transition field of [42],

$$B_z^{\text{isol}\rightarrow\text{comp}} = \frac{\Phi_0}{a_{\text{ch}} a_y} \approx \sqrt{\frac{B_x \Phi_0}{\gamma \lambda_\parallel^2}} \Bigg/ \ln\left[ C\left(\frac{\gamma d}{\lambda_\parallel}\right)^2 \left(\frac{\Phi_0}{\gamma d^2 B_x}\right)^{\frac{1}{2}} \right]. \qquad (4.2.13)$$

Note that the result for $a_{\text{ch}}$ is parametrically the same as that for the lattice spacing, $a_\Delta$, at the first transition to the composite chain+lattice state. The two transitions are only separated by the smaller prefactor for making a dislocation in the full lattice, than for adding a PV stack to an isolated chain. Also, it is an important feature that both transitions are continuous, with the density of PV stacks between chains going continuously to zero at the lower transition, and the density difference between chain and lattice going to zero at the upper transition.

An interesting and important effect within the chain was noticed by Buzdin and Baladie [43]: The individual crossing events on different PV stacks attract each other. The reason for this attraction can be traced to the magnetic interactions between pancakes. Two pancakes in the same layer have a repulsive interaction energy that depends logarithmically on the separation [24-26]. Therefore, when the pancakes are displaced from their stacks at a crossing event, there is an energy contribution equivalent to the interaction between vortex-anti-vortex dipoles, which is simply two derivatives of the logarithmic potential. The same-layer interaction dominates by a factor of $\lambda_\parallel / d$ over the inter-layer interactions, and is given by,

$$E_{n,n}(x) \approx -\frac{\Phi_0^2 d}{8\pi^2 \lambda_\parallel^2} \frac{u_n^2}{x^2}, \qquad (4.2.14)$$

where $u_n$ is the PV displacement in the $n$th layer due to the JV currents. This leads to a long-range attractive contribution to the interaction energy (per unit length) between PV stacks on the same chain,

$$\varepsilon_{\text{att}}(x) \approx -\frac{\Phi_0^2}{\lambda_\parallel^2} \frac{d}{a_z} \frac{u_1^2}{x^2}, \qquad (4.2.15)$$

which must be added to the usual repulsive energy between straight PV stacks. Here, $u_1$ is the displacement of PVs closest to the JV cores, and $a_z$ is the JV separation along the JV stack.

Because of the smaller prefactor but longer range of the attractive term, there will always be a minimum in the PV stack interaction energy along a chain (verified numerically [44]). This has the important consequence that low-density chains are unstable to the attraction. Immediately above the lower critical field for the out-of-plane component, $H_{c1}^z$, the PV stacks should appear with a finite optimal separation $a_{\text{ch}}^*$. In order to have flux densities lower than $\Phi_0/a_{\text{ch}}^*$, there will either be a fraction of JV stacks without chains, or a phase separation along the JV stacks of segments of clustered PV stacks. The optimal separation $a_{ch}^*$ can be estimated by comparing the repulsive force between PV stacks to the attraction between crossing events, with the result (valid for $a_{ch}^* \gg \lambda_\parallel$) [43],

$$a_{ch}^* \approx \lambda_\parallel \ln\left[\left(\frac{\gamma d}{\lambda_\parallel}\right)^2 \left(\frac{\Phi_0}{\gamma d^2 B_x}\right)^{1/2}\right]. \qquad (4.2.16)$$

Again this is parametrically the same result as for the critical separation at the two transitions mentioned above. However, numerical calculations have shown that $a_{ch}^*$ is always smaller than the value of $a_{ch}$ at which PV stacks are ejected from the chain to form the composite state.

One of us has raised the possibility of another, more complex, configuration intermediate between the isolated chain state and the composite chain+lattice state: the buckled chain [44]. When the density of a chain is large enough that the repulsive interactions between PV stacks dominates over the pinning energy to the JV stacks, the excess PV stacks will be ejected from the chain. However, beyond the range $\lambda_\parallel$ of the repulsion, the PV stacks are still attracted to the centre of the JV stacks with a potential of the form [35] (see figure 4.2.2).

$$\varepsilon_{pin}(y) \approx -\frac{\Phi_0^2}{16\pi\gamma a_z y \ln\left(y^2/\gamma d\lambda_\parallel\right)}. \qquad (4.2.17)$$

Therefore, there will be an optimal distance from the centre that a given PV stack is pushed to, and this implies that the repulsion versus attraction can be optimised by the chain buckling, for instance with a profile $y_i = (-1)^i y$, where $y_i$ is the displacement of the $i$th PV stack. The transition from a straight isolated chain to a buckled chain should be first order, while that from a buckled chain to the composite state is continuous.

The interaction energy of the PV stack with a JV stack has a maximum slope as a function of the PV stack displacement, $y$ (see figure 4.2.2). This corresponds to a maximum force that the JV stack can exert on the PV stack, or vice versa. This is an important parameter, and it takes the form [35],

$$f_{max} = -\left.\frac{\partial \varepsilon_{pin}}{\partial y}\right|_{y=y_f} \approx \frac{\Phi_0^2}{4\pi^2\gamma^3 a_z d^2 \ln\left(\lambda_\parallel/d\right)}, \qquad (4.2.18)$$

where the critical displacement is $y_f \approx 0.5\gamma d$. If we impose an external current along the chains, this then corresponds to a critical depinning current for the PV stacks of

$$j_c^x \approx \frac{c\Phi_0}{4\pi^2\gamma^3 a_z d^2 \ln\left(\lambda_\parallel/d\right)}, \qquad (4.2.19)$$

whereas if we drive a current along the z axis, the critical current before the JV's are depinned is,

$$j_c^x \approx \frac{c\Phi_0}{4\pi^2\gamma^3 a_{ch}d^2 \ln\left(\lambda_\parallel/d\right)}. \qquad (4.2.20)$$

The above results have been found using the simple cage approach to individual crossing events. This will break down when $u_1$ is of order $\gamma d$, which occurs for not too large $\gamma$, but also when the PV stacks get too close to each other along the chain. For intermediate values of $\gamma$ and/or high chain densities, a numerical approach is needed to cope with the non-linear equations that describe the phase profile of the crossing configuration. This has been done recently for a single chain by Koshelev [35,45,37] with the following results: For an intermediate range of anisotropy and chain density there is an intermediate structure between the crossing chain and the tilted chain. This is a strongly modulated tilted chain, which can be thought of as straight PV stacks with regular "soliton" deformations to the neighbouring stacks. The three possible chain states are shown schematically in figure 4.2.3. A detailed analysis of the phase boundaries between these different states has recently been carried out by Koshelev [37]. It can be broadly summarized as follows: At high anisotropy, as the out of plane field is turned on, the PVs first enter as stacks, crossing the JVs. Due to the attractive interaction between crossing events, the PV stacks enter sharply with a finite density. On increasing the density further, the effective magnetic trap for the pancakes is weakened (see e.g., [46]) which increases the distortion at the crossing events until the modulated chain is formed. Note that this is a smooth transformation rather than a distinct phase transition. On further increasing the density of PVs, the modulated chain undergoes a continuous phase transition to the straight tilted chain. In contrast, for small enough, or moderate, anisotropies, only the straight tilted chain is seen, as described in section 3.1.

A quite nontrivial behaviour is seen at intermediate anisotropy, and is only revealed with numerical solution of this non-linear problem: For $1.5\,\lambda_\parallel/d < \gamma < 2\,\lambda_\parallel/d$, the tilted chain is seen for low PV densities, as for smaller values of $\gamma$. This is because the energy of a kink along a JV is less than the energy per pancake of a PV stack [29]. However, as the density of PVs is increased there is a first order transition to the crossing chain. This transition has a large jump in density, which can be attributed to the attractive interaction between crossing events, as opposed to the repulsive interaction between kinks along a JV (see, e.g., [8]). On increasing the PV density further in this intermediate regime, the same sequence of transformations to modulated chain and again a straight tilted chain occurs as in the high anisotropy case. Such a re-entrant transition to tilted chain at low PV density was proposed earlier by one of us [44] as a reasonable explanation of the transition seen in [47], where the isolated chain observed from above the crystal sharply changes to a lower density stripe on decreasing the out of plane field (note that the modulations in the field distribution of the tilted chain will be much smaller than is the case for the crossing chain). However, the perturbative calculation performed in [44] contained an error: the correct calculation should not find this re-entrant transition to the tilted chain perturbatively. Even so, the full numerical solution of Koshelev has confirmed the existence of this transition.

To summarize this subsection, we have shown that the crossing lattices exist for high enough anisotropy when the magnetic field is tilted between the $z$ axis and the layer direction. The PV stacks are distorted when they pass close to the JV core. This is the mechanism of an attractive interaction between PV stacks and the centres of the JV stacks that make up the JV lattice. An isolated crossing event is only stable to this distortion for $\gamma > \gamma_c \approx 1.45 \lambda_\parallel / d$. The phase boundary between the crossing lattices and a more conventional tilted lattice is complex, but a simple argument shows that the tilted lattice always exists for small enough field angles from the z axis. In contrast, both possibilities may exist at very small angles from the layer direction, depending on which is largest: the energy of a kink along a JV or the energy per pancake of a PV stack. For low enough $B_z < \Phi_0 / \lambda_\parallel^2$ there is a phase separation between the regular vortex lattice and chains of higher density PV stacks above the JV stack centres. A further decrease of $B_z$ results in a transition to the isolated chain state. The energy profile of the attraction of PV stacks to the JV stacks has the following consequences: a buckled chain is possible that optimises this attraction to the repulsion between nearby PV stacks, and also there is a maximum "pinning" force holding the PV stack to the JV stack. Very low density crossing chains are unstable as there is a long range attraction between crossing events. Finally, at intermediate anisotropy a complex series of transformations of an isolated chain on increasing the PV density has been discovered with numerical solutions of the Lawrence-Doniach model: For $1.5 \lambda_\parallel / d < \gamma < 2 \lambda_\parallel / d$ the PVs first enter the chain as kinks in a straight tilted lattice. There is then a sharp first-order transition to the crossing chain, which smoothly transforms to a modulated tilted chain until at a critical density there is a transition back to the tilted chain. For lower anisotropies than this range the isolated chain is always straight and tilted. For higher anisotropies than this intermediate range, the same sequence of transformations is seen, except that there is no low-density tilted chain.

Finally, it should be noticed that the full phase diagram of the three dimensional problem of vortices in tilted fields in a layered superconductor has not yet been fully described theoretically: the influence of neighbouring chains, or the intervening PV lattice, may change some features. Also, the experiments shown in the next section show strong effects from pinning disorder and thermal fluctuations which still demand to be modelled theoretically.

## 4.3  Experimental Observations of Interacting Crossing Lattices

The first examples of vortex chains in high temperature superconductors were actually observed in highly anisotropic BSCCO single crystals by Bitter decoration [32]. Fig. 4.3.1 shows a micrograph of an as-grown $Bi_{2.1}Sr_{1.9}Ca_{0.9}Cu_2O_{8+\delta}$ crystal ($T_c$=88.5 K) which had been decorated after field-cooling to low temperature in a total field of 35 Oe applied at an angle of $70^o$ to the crystalline c-axis. The image is characterised by high density vortex chains separated by regions of quite well-ordered Abrikosov vortex (AV) lattice. Although the chains are rather disordered they lie predominantly along the direction of the projection of the applied field in the a-b plane. Incommensurability between the chains and the adjacent matrix leads to dislocations in the lattice and the formation of several 5-fold and 7-fold coordinated vortices. At the time that [32] was published an understanding of interacting crossing lattices did not yet exist, and an interpretation was attempted in terms of vortex attraction in moderately anisotropic superconductors [13,48]. These authors made the important observation that the chain spacing at constant tilt angle scaled as the expected AV lattice parameter for the total applied field ($a_y \propto (\Phi_0 / B)^{1/2}$). This is easily understood within the crossing lattices picture since the in-plane field is a constant fraction of the total field at a fixed angle. Within a contemporary interpretation the average chain spacing of $a_y = 9.5\mu m$ in Fig. 4.3.1 reflects the spacing of JV stacks, and an analysis based on anisotropic London theory yields an anisotropy parameter of $\gamma \cong 165$. This relatively low value of anisotropy may explain why the chain state was only observed at tilt angles in the range $60^o$-$85^o$ in these experiments. Although the chains in Fig. 4.3.1 are quite disordered, more recent Bitter decoration images [49] on high quality as-grown BSCCO single crystals ($T_c$=92-93 K) in tilted fields reveal exceptionally ordered chains. Fig. 4.3.2 shows an example of one of these images after decoration at 10K in a field of 20 Oe applied at $70^o$ away from the crystalline c-axis. The almost perfectly straight, periodically spaced chains in this image provide very strong support to an interpretation in terms of 'decorated' JVs. Due to the attractive interaction between JVs and PVs the pancake stacks can be thought of as 'decorating' the underlying JV lattice, by analogy with the Bitter decoration technique itself. In this way the location of the JV stacks, which are extremely hard to 'see' by any other technique, can be visualised indirectly by introducing a small density of PVs. In Fig. 4.3.2, the average chain spacing of $a_y = 18.5$ µm indicates a large crystalline anisotropy of $\gamma \cong 360$. In [49] it was recognised that chain formation in BSCCO could not be explained in terms of theories developed for moderately anisotropic materials, and understood that vortices with different orientations were important. However, it was not until Koshelev demonstrated the explicit coupling mechanism between JV and PV stacks in the crossing lattices phase in 1999 [34] that further progress could be made in understanding them.

A couple of years later interacting crossing lattices were observed in very thin 400nm thick BSCCO crystals by Lorentz microscopy [50]. Note that the thickness is of the order of the in-plane penetration depth $\lambda_\parallel$, which sets the range of attraction between the pancake vortices that make up the stack of an Abrikosov flux line. Fig. 4.3.3 shows a Lorentz micrograph of a BSCCO crystal at 50K with a field of 58Oe applied at $80^o$ away from the crystalline c-axis. The characteristic high density vortex chains aligned roughly along the in-plane field direction and embedded in a well-ordered AV lattice are again observed. The mean chain spacing of about 7.8µm in this image corresponds to a fairly low value of anisotropy $\gamma \cong 195$. The chains indicated by white arrows also appear to kink sharply at the location of a surface step (black arrows). Remarkably it was found that vortex resolution along the chains (but not in the AV matrix) was lost at a temperature, $T_d$, well below the critical temperature of the sample. This is illustrated in Fig. 4.3.4 which shows a sample imaged at 70K in a field of 69Oe applied at $80^o$ to the c-axis. The vortices in the AV matrix are all well resolved, while those in the middle chain have partially disappeared (except for those labelled a and b). The vortices in the upper and lower chains are completely unresolved. The authors attribute this blurring phenomenon to oscillations of vortices in the chains which are too fast to be resolved with the available video imaging rate. In a triangular lattice all vortices are commensurate with the underlying potential landscape, and there is no reason why one group of fluxons should start to move with respect to the others. As was pointed out earlier, however, the chain vortices are incommensurate with the potential distribution imposed by the surrounding AV matrix, and these should be able to move longitudinally at relatively low temperatures [51]. As the temperature is increased a few locally unstable vortices will start to vibrate first which gradually evolves into the oscillation of an entire chain at higher temperatures. The validity of this basic scenario has been confirmed in molecular dynamics simulations [52]. Since the restoring force between the chain vortices will depend on their separation, the disappearance temperature, $T_d$, should depend on the perpendicular component of magnetic field and this is illustrated in Fig. 4.3.5 for a tilt angle of $80^o$. The origins of the negative slope of Fig. 4.3.5 are understood to be two-fold, being due to a lowering of the incommensurate potential at higher fields as well as an increase in the rigidity of the denser vortex chain which helps it to slide freely as a whole. An alternative explanation for the loss of resolution within the vortex chains is that they have undergone a re-entrant transition back to a tilted chain as discussed theoretically in the previous section.

In an ingenious setup, aimed at finding a quantitative measure of the crossing energy, Tamegai *et al*. have introduced parallel straight grooves by Ar-ion milling on to the surface of a BSCCO crystal [53]. A PV stack that sits on one of these grooves has an energy gain roughly equal to the line energy of the now missing pancakes. By observing, with Bitter decoration, when higher density chains appear above the grooves or above the JV's (or both), an estimate was made for the energy gain at a crossing event. The experimental result was around six times higher than the theoretical estimate of (4.2.4), though a more detailed theoretical analysis should be carried out here.

**4.4  Experimental Observation of Isolated Vortex Chains in Interacting Crossing Lattices**

A systematic study of the evolution of the composite chains/lattice vortex phase as the tilt angle was progressively increased towards 90° has been made by scanning Hall probe microscopy (SHPM) [47,54].  The SHPM used was a modified commercial low-temperature scanning tunnelling microscope (STM) in which the usual tunnelling tip had been replaced by a microfabricated GaAs/AlGaAs heterostructure chip containing a sub-micron Hall probe and an integrated STM tip.  After approaching this tip into tunnelling contact the sample was retracted about 100nm, and the sensor rapidly scanned over the sample with the active Hall cross typically 300-400nm above the BSCCO surface [55].  The local Hall voltage was then used to build a 2D map of the surface stray fields.  Fig 4.4.1 shows a set of scanning Hall probe images of the magnetic induction just above the a-b face of an as-grown BSCCO single crystal ($T_c$=90.5K) as the tilt angle, θ, between the applied field and the c-axis is progressively increased.  In these experiments two separate sets of coils were used to generate in-plane and perpendicular magnetic fields, allowing much greater control over the crossing lattice parameters.  When θ=0 (Fig. 4.4.1(a)) a well-ordered hexagonal PV lattice is observed in a field of 12Oe at 81K.  However, as the in-plane field, $H_\parallel$, is increased from zero the composite chains/lattice phase readily forms, as illustrated in Figs. 4.4.1(b),(c), and the average chain spacing corresponds to a rather large value of anisotropy γ≅650.  As has been observed previously the chains are approximately aligned along the direction of the in-plane field (indicated by an arrow) and the vortex separation within the chains is smaller than that of the Abrikosov lattice.  Examples of 7-fold (A) and 5-fold (B) PV rings are also indicated in Fig. 4.4.1(c) which again reflect the incommensurability between the high density chains and the surrounding AV matrix.  Individual PV stacks within the chains are relatively indistinct in these images, but are just resolved with the scanning Hall probe.  This is puzzling as these measurements were taken well above the disappearance temperature established by Lorentz microscopy in reference [49].  The apparent disagreement may be related to differences between the two imaging techniques or, more plausibly, because of the much thinner crystals used in the Lorentz microscopy experiments, which could allow for thermal jumping of the whole flux line.  As the c-axis component of the applied field, $H_z$, is decreased below the "ordering" field [56], quenched disorder destroys the 6-fold symmetry of the Abrikosov domains as shown in Fig. 4.4.1(d).  However, as the tilt angle is increased still further, a new phase transition to another ordered state is observed where all the PV stacks condense onto the vortex chains and the domains of AV lattice disappear.  Figure 4.4.1(e) illustrates this 1D vortex chain configuration, and it is evident that the PVs have assumed the projected quasi-one dimensional symmetry of the underlying JV lattice.

As the field is tilted even further towards the *a-b* plane the PV chains become unstable and a further transition occurs to a state formed of very low flux density, perfectly straight flux stripes which lie along the same direction as the original chains, but with no resolved PV stacks within them (Fig. 4.4.1(f)). It is believed that the latter state of vortex matter corresponds to a re-entrant phase transition to a uniformly tilted vortex lattice. This was first suggested by one of us [44], where numerical evidence was found for such a transition in a single chain. More recently, Koshelev has pointed out an error in this calculation, yet found, within a more restricted range of parameters, that this transition back to a tilted chain does occur on decreasing the PV density [37]. Fig. 4.4.2 shows a few frames from a movie clip which illustrates the changes that occur as the re-entrance phase boundary is crossed in both directions at 83K by cycling $H_z$ slowly at fixed $H_\parallel$=38Oe. At very small c-axis fields the tilted chain state is only visible as a *very* low flux density, structureless flux stripe ($H_z$ = 2Oe and 3Oe). The magnetic induction associated with the stripe grows with increasing $H_z$, as the density of tilted vortices increases. However, as $H_z$ is increased further small regions of the chains begin to transform into segments of interacting crossing lattice, characterised by well-resolved PV stacks. We presume that these stacks cluster together due to their long-range electromagnetic attraction as predicted by Buzdin *et al*. [43]. For quite a broad range of applied fields regions of interacting PV stacks coexist with the tilted vortex phase and have a ten times higher flux density ($H_z$ = 4Oe and 4.4Oe). This implies a first order transition, and is in fair quantitative agreement with recent calculations [37]. If $H_z$ is reduced again the PV stack segments shrink ($H_z$ = 0Oe) and eventually all flux leaves the sample ($H_z$ = –2Oe). If the field is decreased further a tilted phase of opposite sign is observed ($H_z$ = -3Oe) and the nucleation of 'black' interacting PV stacks occurs ($H_z$ = - 4Oe). It is interesting to note that interacting chains of PV stacks are rarely straight indicating that they can readily accommodate to regions of quenched disorder to maximise their pinning energy. In contrast the re-entrant tilted phase exists in perfectly straight stripes, which may be explained by the fact that the pancake vortices that make the kinks in a tilted flux line are more mobile, and thermally hop over the local pinning barriers.

SHPM measurements have been used to provide a quantitative test of Koshelev's theory of interacting crossing lattices [34]. The left hand inset of Fig. 4.4.3 shows an SHPM image of the magnetic induction just above the a-b face of an as-grown BSCCO single crystal ($T_c$=90K) at 81K with the magnetic field at an angle of 89° to the *c*-axis ($H_z$ =0.7Oe, $H_\parallel$=49.5Oe). Assuming that these chains correspond to the 'decoration' of the underlying JV lattice by PV stacks in the interacting crossing lattices ground state we expect that the chain spacing will be independent of $H_z$ (if $H_z$ is small) and will directly mirror the in-plane JV lattice spacing. Within anisotropic London theory the lateral separation of adjacent JV stacks is given by (4.1.3). Fig. 4.4.3 illustrates how the chain separation, $a_y$, (at various small values of $H_z$) at four different temperatures in the range 77-85K falls close to a single universal line when plotted versus $1/\sqrt{H_\parallel}$. A linear regression fit to the 81K data (black line) yields an anisotropy parameter $\gamma$ = 640±25. The right hand inset of Fig. 4.4.3 shows that the temperature-dependence of the measured anisotropy at $H_{||}$=38Oe is extremely weak in this temperature range. At very high in-plane fields the layered nature of BSCCO breaks the continuous London approximation and this is reflected in the fact that the linear fit to the data intersects the y-axis below the origin at −1.3±0.3µm. Computing corrections to the Josephson vortex interactions due to the layered structure and the corresponding deformation of the lattice, the following expression was obtained [37]

$$a_y \approx a_{y0}\left(1 - \frac{9\sqrt{3}\gamma d^2 H_{||}}{32\Phi_0}\right) \qquad (4.4.1)$$

where $a_{y0}$ is the uncorrected chain spacing and d is the separation of the $CuO_2$ layers in the crystal. This behavior is plotted as the gray line in Fig. 4.4.3 and, although it qualitatively describes the trends observed, it underestimates the experimental deviations.

Quenched disorder in BSCCO single crystals shows a pronounced anisotropy, with the consequence that even for fields along the *c* axis one frequently observes the formation of pinned PV chains along the crystallographic *a*-axis [57]. Therefore, care must be taken to distinguish these chains that are due to pinning to the crystal inhomogeneity from the chains created from an attraction to a coexisting JV lattice. Even so, one can take advantage of the pinning phenomenon to *directly* observe the relaxation of the PV stack structure in the presence of JV currents, which is what gives rise to the attractive PV/JV interaction in the crossing lattices regime. Fig. 4.4.4(a) shows an image of two weakly pinned (*a*-axis) chains after field-cooling a BSCCO sample to 83K in $H_z$=1.8Oe ($H_{//}$=0) followed by the application of a small in-plane field $H_{//}$=11Oe along the *a*-axis. Fig. 4.4.4(b) shows exactly the same vortex configuration after the in-plane field was increased to $H_{//}$=16.5Oe, at which point a pronounced smearing of the top right-hand vortex chain was observed while all other flux structures remained unchanged. The smearing reflects small PV displacements which are driven by the currents due to the JV stack which is now aligned along it. As the in-plane field increases, the separation of Josephson vortices that cross a PV stack decreases, leading to more displacements and a larger smearing. Fig. 4.4.4(c) shows linescans along the relevant chain before and after the increase of in-plane field, clearly illustrating how the field modulation along the chain falls by about a third for $H_{//}$=16.5Oe. The *n*th PV above a given JV core is predicted [34] to experience a lateral displacement given by (4.2.3). The solid lines through the data represent fits generated by entering $u_n$ into a pancake vortex model due to Clem [58] which yields the following expression for the magnetic induction at a height h above the center of an isolated chain of PV stacks.

$$B_z(x,h) = \frac{d\Phi_0}{2\pi\lambda_{//}^2 a_{ch}}\sum_n\sum_{G_i}\int_{-\infty}^{\infty}\frac{\exp(-\sqrt{G_i^2+q_y^2+\lambda_{//}^{-2}}.n.d).\exp(-\sqrt{G_i^2+q_y^2}.h).\cos(G_i.(x-u_n))}{\sqrt{G_i^2+q_y^2+\lambda_{//}^{-2}}+\sqrt{G_i^2+q_y^2}}.dq_y$$

$$(4.4.2)$$

Here $G_i = 2\pi i / a_{ch}$ are the reciprocal lattice vectors of the chain and $n$ is summed over all $CuO_2$ bilayers down through the thickness of the sample. Care was taken to include the effects of surface barriers on the JV lattice, the depth of the first JV beneath the sample surface being given by

$$z_1 = a_z / 4 \left( 1 + \sqrt{1/3 + (2/\sqrt{3}\pi) \ln\left(2.86\Phi_0 / 2\pi\gamma d^2 B_x\right)} \right)$$ [59], where $a_z$, given by (4.1.2), is assumed

constant below the first JV. Equation (4.4.2), together with the formula for $u_n$, was solved numerically assuming $\lambda_\parallel(83K) = 450nm$, $r_w = 270nm$, $d = 1.5nm$, $a_z = 48nm$, $a_{ch} = 2.35\mu m$ and $\gamma = 640$. The scan height, $h$, was treated as a fitting parameter for the $H_\parallel = 11Oe$ data, and the $H_\parallel = 16.5Oe$ trace was then calculated with effectively no free parameters. With $h = 650nm$ and $u_n = 0$ an excellent fit was achieved to the linescan from Fig. 4.4.4(a) for the non-interacting chain. In addition the degree of broadening due to the PV displacements shown in Fig. 4.4.4(b) is also well reproduced when $u_n \neq 0$, lending excellent support to the underlying theory.

Earlier measurements of the composite chains/lattice state have shown that quenched disorder has a pronounced effect on interacting crossing lattices, leading in many cases to significant disordering of the chains. Remarkably, it is found that the JV/PV interaction is sufficiently strong that JV stacks become *indirectly* pinned at the location of pinned PV stacks since they can interact with a higher density of PVs in these regions. Indirect pinning of JVs is more pronounced at high PV densities due to the increased intrachain PV repulsion. The crooked bright line on the right hand side of Fig. 4.4.5(a) represents a strongly 'kinked' stack of JVs due to indirect pinning by PVs along the crystalline a-axis (in the direction of the arrow shown) down a 9μm long segment in its centre at 85K ($H_\parallel = 35.8Oe$, $H_z = 8.8Oe$). Similar, but less dramatic behavior is observed in Fig. 4.4.5(b) at 83K with a lower PV density ($H_\parallel = 35.8Oe$, $H_z = 4.4Oe$). The problem of indirect pinning for Josephson vortices at a finite angle away from the symmetry axis of the pancake-vortex lattice is analogous to the deformation of tilted vortices near columnar defects [60]. A simple model that assumes the pinned Josephson vortex to be an elastic string within a "cage" potential (that represents the interactions with the rest of the Josephson vortex lattice) allows one to estimate the length of the pinned segment as a function of the *indirect* pinning potential per unit length, $U_p$, the cage potential $K = \Phi_0 B / 4\pi\gamma^2\lambda_\parallel^2$, and the JV line tension $\sigma \cong \Phi_0^2 / ((4\pi)^2\gamma\lambda_\parallel^2)$, as well as the angle, θ, between the in-plane field and the strong pinning axis. The model assumes that the JV is exactly pinned to the center of a PV chain along the length $w$, while beyond the pinning segment it relaxes as an elastic string towards the center of the cage, which seems an accurate model of the feature in figure 4.4.5(a). Balancing the pinning, elastic and cage energies one finds that the length of an indirectly pinned JV stack is given by

$$w = \sqrt{\frac{8.U_p}{K.\sin^2\theta.\cos\theta}} - \sqrt{\frac{4.\sigma}{K.\cos^2\theta}} \qquad (4.4.3)$$

Assuming that the pinning potential can be approximated by $U_p = E_\times \Delta(1/a_{ch})$, where $E_\times$ is given in (4.2.4) and $\Delta(1/a_{ch})$ is the difference in PV chain density between the strongly pinned region and elsewhere, quantitative agreement can be obtained between the pinned length calculated from eqn. (4.3.3) and the measured one in Fig. 4.4.5(a) using the same parameters as before and $\Delta(1/a_{ch}) \cong 1\,\mathrm{PV\,stack/\mu m}$.

Isolated vortex chains were soon observed by several other techniques, the first being magneto-optical imaging (MO) [61,62]. In these MO experiments a magnetic garnet indicator was pressed into close proximity with the sample and illuminated with polarised light. The light experiences a small Faraday rotation in its polarisation as it passes through the indicator which depends linearly on the local stray field. These very small rotations can be detected using crossed polarisers for incident and reflected light, and captured with a digital camera to build an image of the magnetic field at the sample surface. Fig. 4.4.6 shows several magneto-optical images of a BSCCO single crystal ($T_c$~90K) at 82K in a perpendicular field of 2Oe and various in-plane fields (as indicated) [63]. For in-plane fields larger than 2Oe quite good images of PV chains could be obtained in this region of the sample. Although these lie approximately along the in-plane field direction, there is considerable disorder at low fields when the inter-chain interactions are weak. One advantage of the MO technique is that one can image a very large area at one time, as illustrated in Fig. 4.4.7 at 80K for a region of sample almost 1mm wide ($H_\parallel$=36 Oe, and $H_z$= 2 Oe). While chains are visible over much of the sample it is clear that they are heavily defected with at least one very bright 'flux concentrator' visible. Moreover, a careful analysis of the mean chain spacing in different regions of the sample revealed that it could vary dramatically from place to place. Fig. 4.4.8 shows two small images from distant parts of the crystal which were imaged at the same time under identical conditions ($H_\parallel$=21 Oe, $H_z$=1.1 Oe, $T$=85K). It is evident that the chain spacing can vary by up to a factor of two across the sample which can presumably be attributed, in part, to relatively large anisotropy inhomogeneities due to structural and compositional variations. These authors also observed steps in plots of the chain spacing as a function of in-plane field. These were not an imaging artefact and were attributed to non-equilibrium behaviour of the JV lattice. When the in-plane field is increased JVs have to cross $CuO_2$ planes, a process that is hindered by significant intrinsic pinning. Hence it is likely that the system does not always reach its equilibrium configuration. For $H_z$=0 it is known that there are many degenerate structures for the JV lattice [27], with the spacing between JV stacks varying from $a_{y1} = (\sqrt{3}\gamma\Phi_0/2B_x)^{1/2}$ to $a_{y2} = (\gamma\Phi_0/2\sqrt{3}B_x)^{1/2}$. For finite $H_z$ the structure with $a_{y1}$ is slightly more favourable, as we have assumed elsewhere in this review. However, it is quite possible that other structures may coexist in a non-equilibrium state. This may also account for some of the variation in chain spacing observed across the sample. These observations are in agreement with the results of SHPM imaging experiments where it was found that dithering the in-plane field several times around the desired value leads to much more consistent sets of chain spacing data. Presumably the dithering process helps the vortex system reach its equilibrium state. In addition, magneto-optical measurements at very low perpendicular fields revealed nearby regions where the chain spacing varied by almost exactly a factor of two. This was interpreted as a consequence of the long-range attractive interaction between crossing PV stacks [43] which gives rise to an optimum stack spacing. When the PV stack density is too low to achieve this optimal spacing everywhere the system can lower its energy by only decorating, for example, every second JV stack. Finally, a close examination of the data shown in Fig. 4.4.6, particularly that at low in-plane fields, reveals characteristic defects in the JV lattice. Frequently "forks" appear, which are interpreted as dislocations which transfer JV lines between separate JV stacks, and confirms that PVs are capable of decorating JVs at various depths beneath the surface (as it is not possible for a single JV to split in two).

Lorenz microscopy is one of the few magnetic imaging techniques that is sensitive to the magnetic field distribution *inside* superconducting samples. It is capable, therefore, of distinguishing between tilted vortices, whose direction follows that of the applied field, and crossing lattices which do not. As shown earlier, Lorentz-mode vortex images in the chain state of YBCO elongate as the tilt angle increases, as expected within a tilted lattice model. In contrast Fig 4.4.9 shows a series of images of isolated chains as a function of tilt angle at 50K in a very thin (300-400nm) BSCCO crystal [20]. The in-plane field had been kept constant at 50Oe in these measurements, while the perpendicular field was gradually increased from zero. There is no indication at all that the vortices broaden as the tilt angle is increased (smaller $H_z$) consistent with the fact that one is observing a crossing lattice state here as expected. There is also clear evidence for selective population of adjacent JV stacks in the image at $H_z$=1Oe. This is in agreement with similar magneto-optical observations, and probably also arises due to the long-range attraction between interacting PV stacks.

Isolated vortex chains have also been resolved in high-resolution Bitter decoration images [64]. Fig. 4.4.10 shows an image of the edge of a highly overdoped ($T_c$=77K) BSCCO single crystal which had been field-cooled in an in-plane field of 100Oe. Cleverly, the dome-like profile of the magnetic induction perpendicular to a plate-like sample (due to the geometric surface barrier, see [65]) was used to provide a range of $B_z$ over which the crossing lattices could be observed. To be precise, a perpendicular field of 15Oe was applied when the sample reached a temperature of 30K leading to smooth gradients of $B_z$ near the sample's edge. Decoration was then performed after further cooling the sample to 4K. The induction gradients remain locked into the sample and enable crossing lattice phases to be studied for a range of values of $B_z$ in a single decoration. The edge of the BSCCO crystal is in the lower left corner of Fig. 4.4.10 (bright diagonal stripe), and we see that the shielding currents induced at the sample edges have driven the vortices towards the center of the sample resulting in the expected dome-like distribution of PV stacks. Near the edge of the sample, where the PV density is low, we clearly observe the isolated chain state approximately along the direction of the in-plane field (white arrow). Towards the center of the sample the PV density increases, JV stacks become saturated with interacting stacks and the system crosses over to the composite chains/lattice state. The average chain spacing of 5.2μm corresponds to an anisotropy of $\gamma \cong 180$ for this overdoped crystal, which is close to, but still above the critical value of $\gamma_c = 1.45 \, \lambda_\parallel/d$ below which crossing events are unstable (see section 4.2). In addition, by numerically evaluating the local magnetic induction these authors found that the transition to the chain state takes place at $<B_z> \cong 3.5$G for this value of in-plane field. Isolated chains could also be decorated in optimally doped BSCCO crystals ($T_c$=90K) as shown in Fig. 4.4.11 for $H_\parallel$=100Oe and $H_z$=10Oe. In this sample the mean chain spacing of about 10μm corresponds to a much large anisotropy of $\gamma \cong 560$, and the critical induction at which the transition to the composite chains/lattice state occurs was estimated to be $<B_z> \cong 1.5$ G, qualitatively consistent with the expected reduction for higher anisotropy materials [$(H_z^{cr})^2 / H_\parallel \sim 1/\gamma$, see (4.2.13)]. The spatial positions of the PV stacks along the chain indicated by the dashed white box have been reproduced in the inset at the top of the figure for clarity. Several of the PV stacks near the center of the chain deviate substantially from the expected linear behaviour, and show a pronounced zigzag structure. The authors speculate that this may be the first observation of the buckled state predicted by one of us [44] due to the repulsive interaction between the stacks at high linear densities (i.e. at separations much less than the optimum value calculated in reference [43]). Indeed the estimated mean induction of 1.6G, where possible buckling is observed, compares well with the theoretically estimated threshold field of 1.9G for $H_\parallel$=100Oe, and the buckling amplitudes of ~1μm are also in good agreement with predictions.

A more recent experiment [66] has looked at isolated chains in higher in-plane fields of $H_\parallel$~1000Oe. Here a surprising phenemomen of bundles of chains was seen with Bitter decoration, coexisting with simpler chains. This was explained within the same buckling scenario of [44], but where an attraction between different buckled chains becomes possible, as a PV stack gains more distortion energy by being near two JV stacks. As the in-plane field is increased, larger bundles appear, and there are fewer simple isolated chains, until above a critical field the sample is filled by these dense chains.

Finally, we note that recent detailed measurements of the flux profile in a chain, both with SHPM and with Bitter decoration, show an asymmetry that is well explained by a small tilt of the PV stacks [67]. This is understood to be due to the net torque on each segment of PV stack between neighbouring JVs, due to the Lorentz force from in-plane currents. The effect is enhanced by an increased out-of-equilibrium magnetization.

**4.5 Dynamic Properties of Isolated Vortex Chains in the Crossing Lattices Regime**

One of the most interesting and important experimental properties of the isolated chain state is that the underlying JV stacks represent robust "one-dimensional" traps for PV stacks, dominant over the intrinsic PV pinning, over a broad temperature range (77-88K). This may be anticipated close to $T_c$ because the PV-JV interaction is almost temperature independent [34] while the PV pinning energy decreases rapidly close to the transition temperature. In addition the short-range repulsion between PVs in the same $CuO_2$ plane prevents PV stacks from passing one another along the length of the chain. As a consequence the dynamic properties of the isolated vortex chain state appear to be particularly rich as illustrated by the following two complementary SHPM imaging experiments where one magnetic field component is held constant and the other varied.

Fig. 4.5.1 shows SHPM images of the dragging of a pancake chain by a JV stack at 83K as $H_\parallel$ is reduced at fixed $H_z$=0.5Oe [47]. During this coherent movement the JV stack (J) is able to depin and drag with it two isolated PV stacks (A, B) which were originally pinned on the right hand side of the image. Pancake motion induced by the motion of JV stacks is unusual because these PV displacements are reversible and uniquely determined by the applied field $H_\parallel$, in stark contrast to the customary situation where c-axis magnetic fields and/or in-plane currents determine the forces acting on PVs rather than vortex positions. It has even been shown [47], for example, that it is possible to demagnetise a sample containing trapped PV stacks by turning on a large in-plane field, to drive the system into the 1D vortex chain state, and then slowly removing it. As the JVs are gradually swept out of the sample they drag the PV stacks with them, leaving a fully demagnetised sample. The SHPM image shown in Fig. 4.5.2 illustrates this magnetic 'brush' effect rather graphically. In this case a BSCCO sample had been field-cooled to 77K in $H_z$=10Oe ($H_\parallel$=0), and then $H_z$ had been set to zero leaving a fairly large density of PV stacks trapped at regions of quenched disorder in the crystal. The in-plane field was then swept up to $H_\parallel$=39Oe generating JV stacks which flowed across the sample. In doing so it is clear that they have gathered up many PV stacks in their direction of motion (arrow shown) leaving a region completely free of PV stacks behind them.

It is known that the irreversible magnetisation of BSCCO single crystals can be made reversible by the application of in-plane ac fields [68]. Figs. 4.5.3(a)-(c) show SHPM images of the increase of the pancake density inside a PV chain trapped on a JV stack at 81K as the c-axis field is increased from zero at fixed $H_\parallel$=35Oe [47]. It was observed that the PV mobility along JV stacks appears to be considerably higher than in JV-free regions, and the presence of JVs allows PVs to overcome quenched disorder, effectively depinning them in the direction of the JV. Fig. 4.5.3(d) shows local magnetisation loops ($\mu_o M_{loc}$ versus $H_z$) at constant $H_\parallel$=35Oe, measured with the scanning Hall probe fixed ~1μm above the same region of the sample, which reveal a dramatic suppression of the irreversibility as compared to measurements in the absence of JVs ($H_\parallel = 0$). The presence of $H_\parallel$ also slightly decreases the PV penetration field ($H_p$). At fields just above $H_p$, PV stacks preferentially enter the sample along JVs, where the superposition of Meissner and JV currents at the edges leads to a slight lowering of Bean-Livingston penetration barriers [69], and then enjoy a much higher mobility along the JV chains. PVs only then propagate into the interchain spaces if the chain density becomes saturated, i.e. if $H_z^2 / H_\parallel > H_{tr}(T, H_\parallel)$. Conversely, if $H_z$ is reduced, PVs exit by travelling along JVs to the sample edges. The ability to manipulate the positions and trajectories of PV stacks by controlling the underlying interacting JV lattice opens up new possibilities for conceiving functional vortex devices. Several possible types of these have recently been discussed by Savel'ev *et al*. [70].

Finally, SHPM measurements have shown [54] that fragmentation of PV and JV stacks can occur when 'decorated' stacks of JVs are forced abruptly through a region of disorder which is inhomogeneous along the c-axis. Fig. 4.5.4(a) shows an image of a pinned chain of PV stacks ($H_z$=0, $T$=85K) after the in-plane field was suddenly reduced from $H_\parallel$=36Oe to zero. Note that a 'fractional' pancake vortex stack has now formed in the middle of the chain which 'healed' back to a connected stack after the in-plane field was cycled back up to $H_\parallel$=11Oe (Fig. 4.5.4(b)). Fig. 4.5.4(d) shows linescans across these images in the indicated directions, and it was found that Clem's pancake vortex model [58] described the data well under the assumption that a PV stack had split cleanly, at a depth of ~480nm below the surface, into two segments a lateral distance 2.3μm apart (Fig. 4.5.4(c) and gray curve in Fig 4.5.4(d)). Similarly Fig. 4.5.5(a) illustrates the splitting of a decorated JV stack into two 'forks' at 77K after $H_\parallel$ was abruptly increased from –33Oe to 33Oe ($H_z$=1Oe) [54]. Note that the PV chain density in each fork is approximately the same indicating that the JV stack must have split into two almost equal sections. Fig. 4.5.5(b) shows how the split chain has recombined after the in-plane field was reduced to $H_\parallel$=22Oe in the subsequent scan.

## 5. Concluding remarks

In conclusion, we have described how the extremely high crystalline anisotropy in layered superconductors, in particular high-$T_c$ cuprate superconductors, has a dramatic influence on the vortex structures formed. The effect of the anisotropy is particularly prominent when the applied field is tilted away from a high symmetry axis (e.g. the c-axis perpendicular to the $CuO_2$ planes in high-$T_c$ materials). One of the more surprising consequences is the instability of the well-known triangular Abrikosov vortex lattice with respect to the formation of structures based around chains of flux lines. We have reviewed the experimental observations of chain structures and their theoretical understanding in the two distinctly different cases of moderate and high anisotropy. In the former case an attractive part of the interaction between tilted flux lines leads to chain formation, while in the latter it is related to the interaction of two independent and perpendicular crossing lattices of Josephson and pancake vortices. The latter system of crossing lattices exhibits particularly rich phenomena with many different equilibrium phases depending on the direction and magnitude of the applied magnetic field. In particular, the fact that pancake vortex stacks become attracted to underlying Josephson vortices allows one to visualise the JVs indirectly (by virtue of the excess density of PVs that accumulates along them), something that is very difficult to achieve by any other imaging method. The term 'decoration' has been coined to describe this technique due to the rather direct analogy with the well-known Bitter decoration method. Assuming that the presence of the interacting PVs only represents a weak perturbation, this approach can be used to study the structure of the JV system as well as search for new equilibrium states of interacting PV and JV matter.

There is also an extraordinary dynamic interplay between the two sets of crossing vortices. On the one hand moving JV stacks can drag PV stacks along with them like a vortex 'brush', while on the other the underlying JV stacks represent high mobility channels which guide the transport and penetration of PV stacks. Both these properties have been exploited to completely suppress the irreversible out-of-plane magnetisation in BSCCO single crystals by applying a simultaneous 1kHz ac dithering field in the a-b plane [68]. There are also a number of functional vortex devices that could potentially arise from interacting crossing lattices. Savel'ev *et al.* [70] have proposed a family of vortex pumps, diodes and lenses that are based on the vortex 'brush' phenomenon. It is even possible to envisage an entirely new form of vortex logic based around a single PV stack 'bit' which is manipulated and guided by controlling the underlying JV system. To some degree the problem could also be reversed, and JV stacks moved around by modifying the PV system. Several groups around the world are currently actively trying to demonstrate prototype systems with these underlying operation principles.

Clearly the system of interacting crossing vortex lattices not only provides a unique toolbox for probing the formation of structures with competing interactions, but also has considerable potential applications. It seems likely that there may be several more phases of interacting JVs and PVs which remain to be discovered, and work on potential applications is still in its infancy. This area of research, therefore, looks likely to remain an exciting and dynamic one for some time to come.

## 6. Acknowledgements

We are very grateful to A.E.Koshelev and J.R.Clem for sharing theoretical results with us prior to publication. We would also like to express our thanks to the following people who have granted us permission to reproduce their magnetic images and other data in this review article; J.Karpinski, L.Ya.Vinnikov, T.R.Dinger, P.L.Gammel, D.J.Bishop, A.Tonomura, C.A.Bolle, I.V.Grigorieva, A.N.Grigorenko, V.K.Vlasko-Vlasov, T.Tamegai & T.Tokunaga. One of us (MJWD) would like to thank the Solid State Theory group at Centro Atomico Bariloche for their hospitality during the writing of some of this article.

MJWD was funded by the Swiss National Science Foundation via the national centre of competence in research (MaNEP) and by an EPSRC advanced research fellowship GR/A90725/02. SJB acknowledges financial support from EPSRC in the UK through grant no. GR/R46489/01, and from the ESF VORTEX programme.

**Figures**

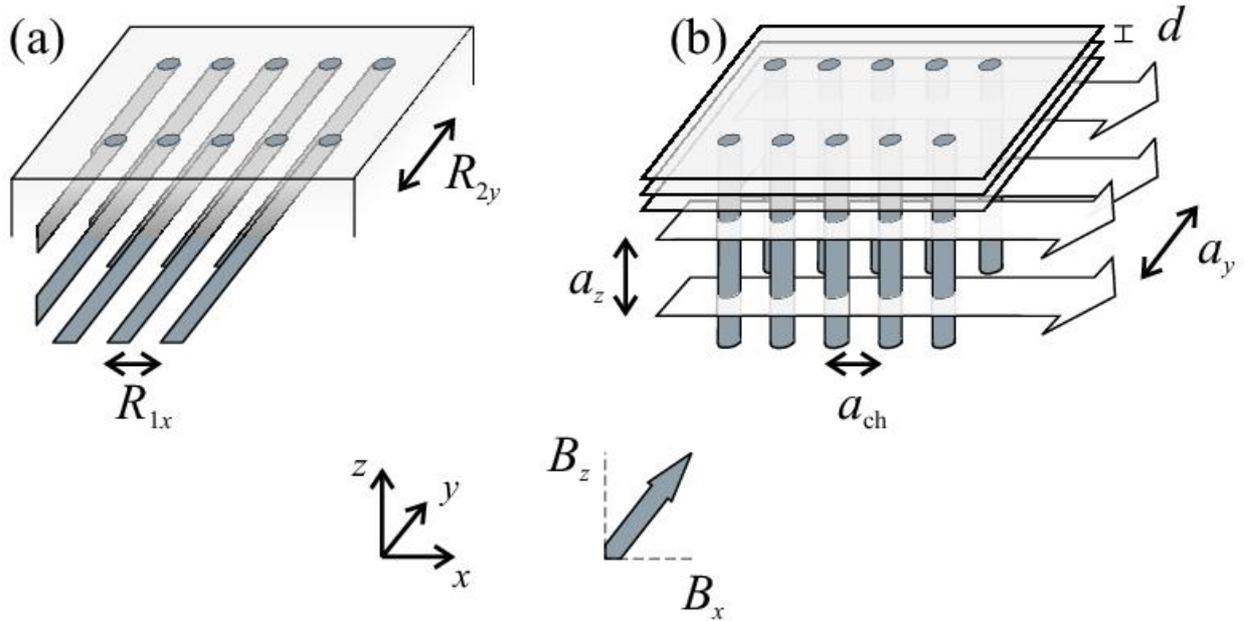

Fig. 2.1  A schematic diagram showing the geometry we consider in this paper: We define the layers of an anisotropic superconductor to lie in the x-y plane, and a tilted magnetic field to lie in the x-z plane.  In (a) we show chains of tilted flux lines, which form a two-dimensional lattice in the x-y plane with lattice vectors $\boldsymbol{R}_{1x}$ and $\boldsymbol{R}_{2y}$.  The crossing lattice state is shown in (b), where chains of pancake-vortex stacks form, due to the attraction to the centres of the in-plane Josephson vortices.  This state only occurs in a highly anisotropic layered superconductor.  The layer separation is $d$.  The distance between neighbouring PV stacks in a chain is $a_{ch}$.  The separation of JV's in a stack is $a_z$ and the distance between chains is $a_y$.

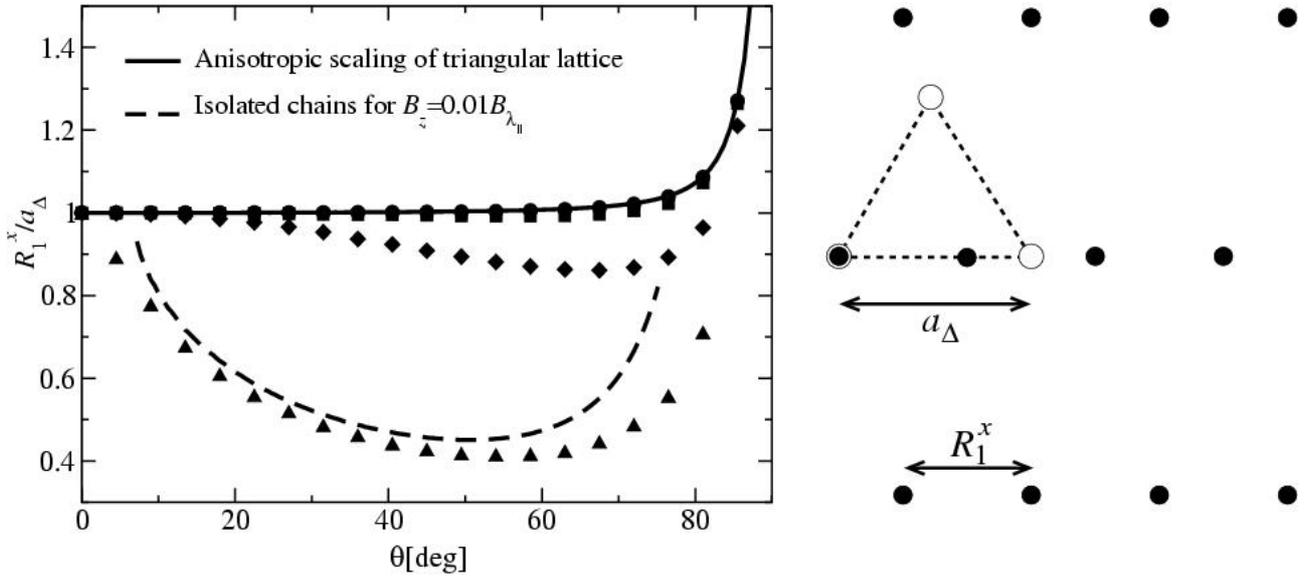

Fig. 3.1.1  Results of a numerical minimization of the tilted lattice within the continuous anisotropic London theory.  Plotted is the lattice parameter in the $x$ direction when the field is in the $x$-$z$ plane, against the field angle from the $z$ axis.  The different numerical results are for $\gamma=10$ and $B_z=0.01$, 0.1, 1 and 10 $B_{\lambda_{\parallel}}$ (triangles, diamonds, squares and circles respectively), where $B_{\lambda_{\parallel}}=\Phi_0 \big/ \lambda_{\parallel}^2$ . The solid line is the result from anisotropic rescaling theory, $R_1^x = a_\Delta \sqrt{1+\gamma^{-2}\tan^2\theta}$ , where $a_\Delta$ is the lattice spacing of an unstretched triangular lattice, as shown in the figure.  The dashed line is the result for $B_z = 0.01 B_{\lambda_{\parallel}}$ when $R_1^x$ is fixed at the optimal value for an isolated chain.

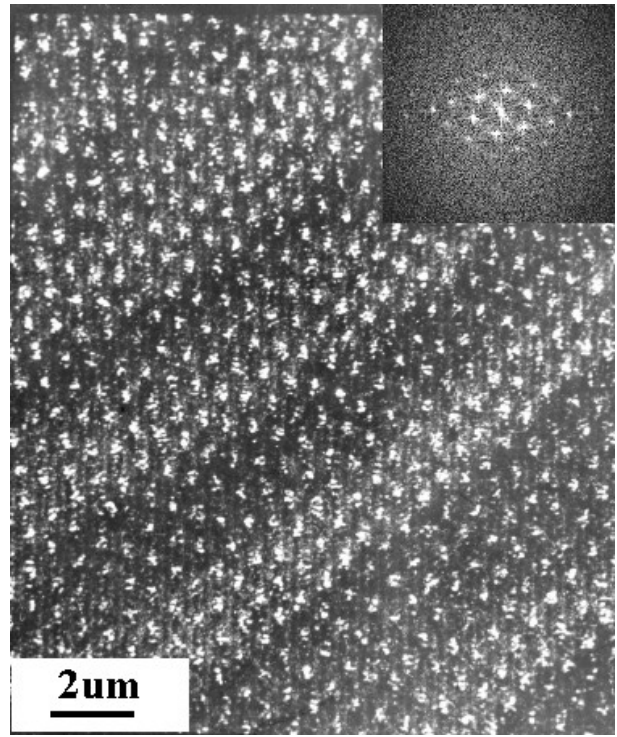

Fig.3.2.1  Micrograph of a YBCO-124 single crystal which had been Bitter decorated after field-cooling to low temperature in an applied  field of $H_z$=41Oe.  A digital Fourier transform of the image is inset in the top right hand corner.  [*Crystal grown by Prof. J.Karpinski ETH, Zürich; Bitter decoration by Prof. L.Ya.Vinnikov, Institute of Solid State Physics RAS, Chernogolovka, Russia*]

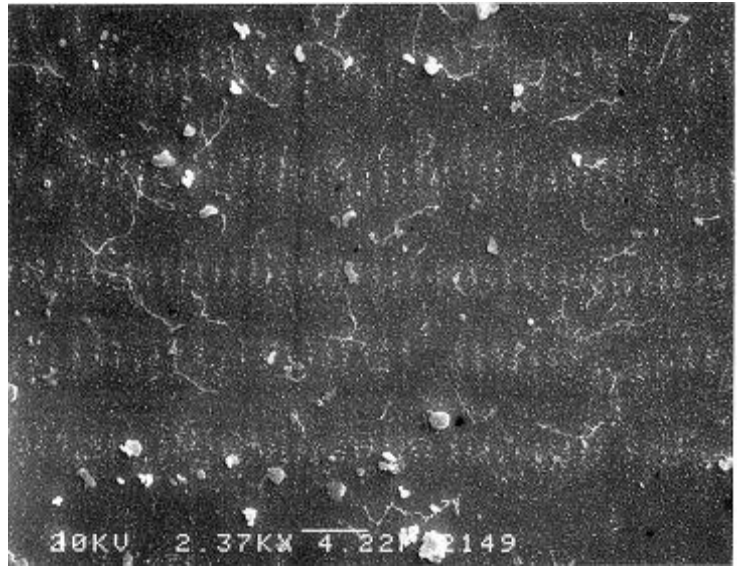

Fig. 3.2.2  Bitter decoration image of a plane perpendicular to the c-axis of a YBCO single crystal after field-cooling to low temperature in $H_\parallel$=4Oe.  The white marker bar is 10μm long.

[*Reproduced with permission from reference [15]. Copyright 1989 American Physical Society*]

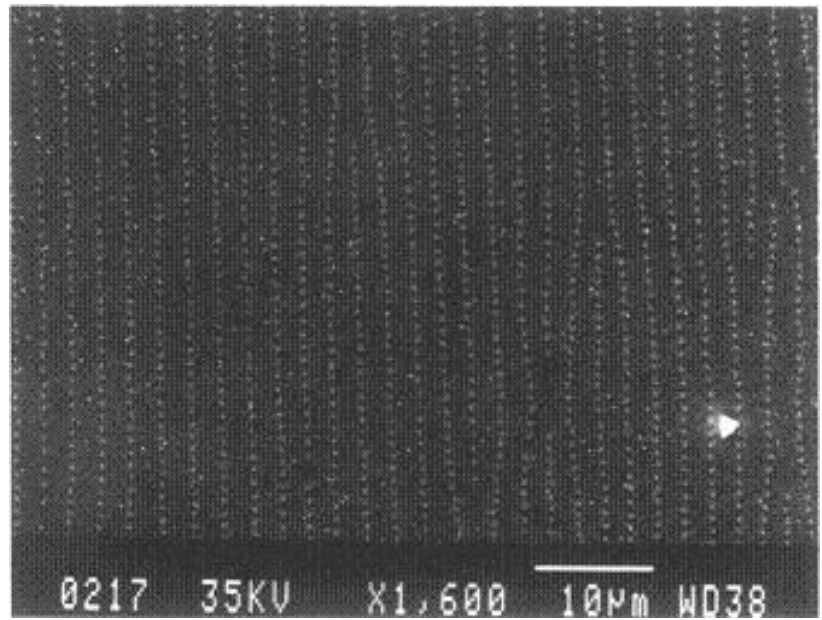

Fig. 3.2.3  Bitter decoration image of the a-b surface of a YBCO crystal after field-cooling to low temperature in a total field of 24.8Oe applied at 70$^o$ from the crystalline c-axis.  [*Reproduced with permission from reference [19]. Copyright 1992 American Physical Society*]

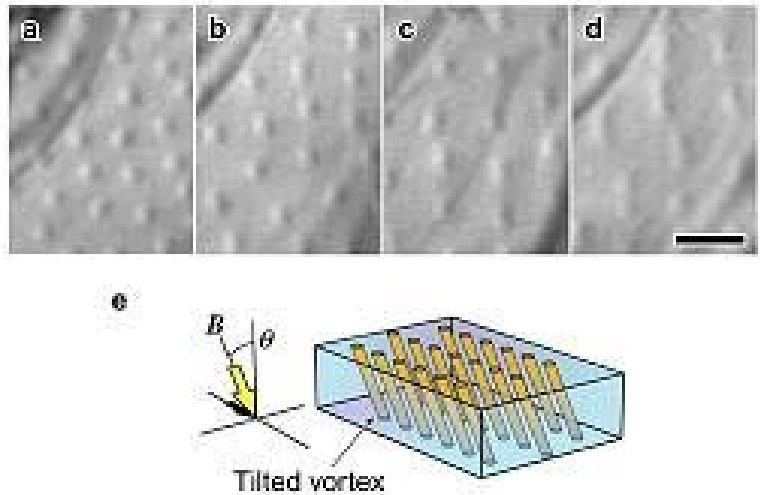

Fig. 3.2.4  Lorentz micrographs of vortices in a very thin YBCO crystal at 30K and $B_z$=3G as a function of tilt angle.  (a) θ=75°, (b) θ=82°, (c) θ=83° and (d) θ=75°.  (e)  Schematic diagram of the tilted vortex lines.  [*Reproduced with permission from reference [20]. Copyright 2002 American Physical Society*]

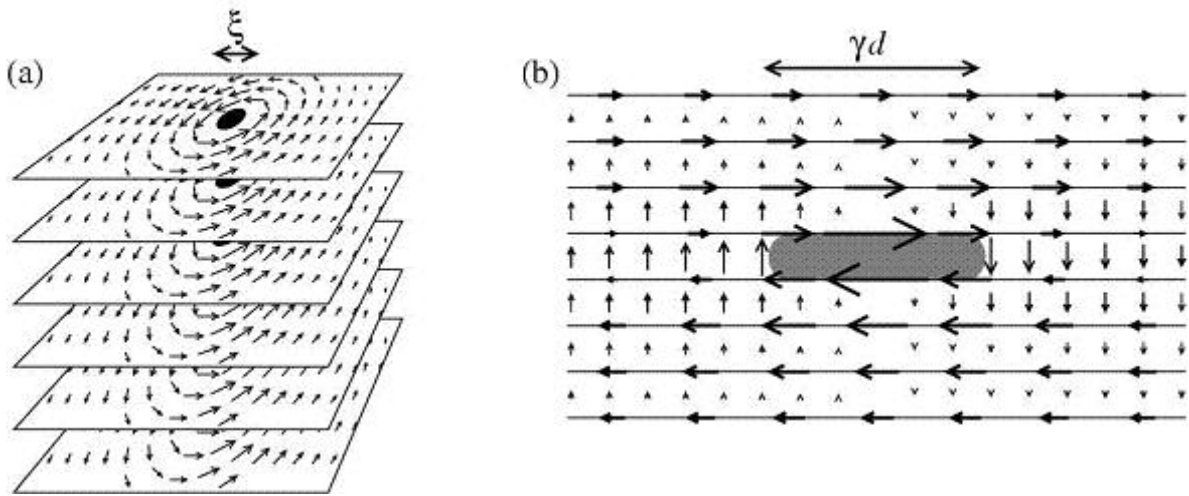

Figure 4.1.1  Schematic diagrams of the current flow in (a) a pancake vortex stack and (b) a Josephson vortex.  In the PV stack the currents are only within the layers, as is the core where the order parameter is suppressed.  In the JV there are also currents from layer to layer, but there is no core suppression as the vortex centre lies between two layers.

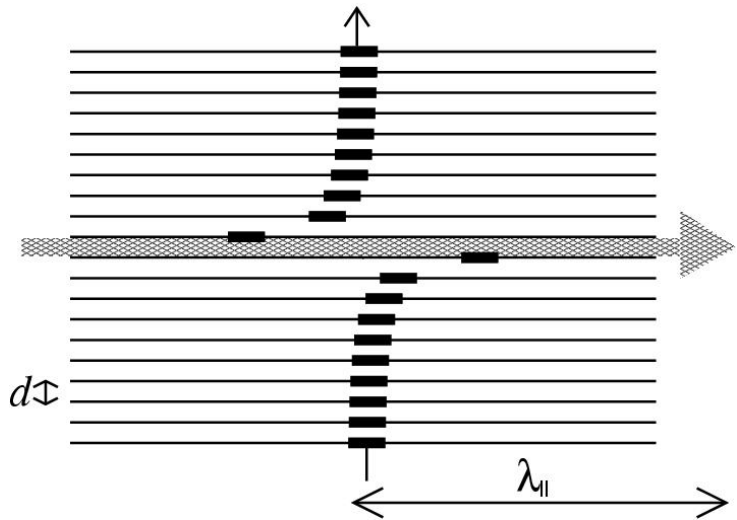

Figure 4.2.1  Pancake vortex configuration when a PV stack crosses through the centre of a JV. The Lorentz force on the pancakes from the JV currents pulls them away from the stack centre, until this is balanced by their attraction to the remainder of the stack.  The displacement is equal and opposite above and below the JV, and is largest for the two pancakes immediately adjacent to the JV core.  This configuration is (meta)stable for large enough anisotropy, $\gamma > \gamma_c \approx 1.45\,\lambda_\parallel / d$.

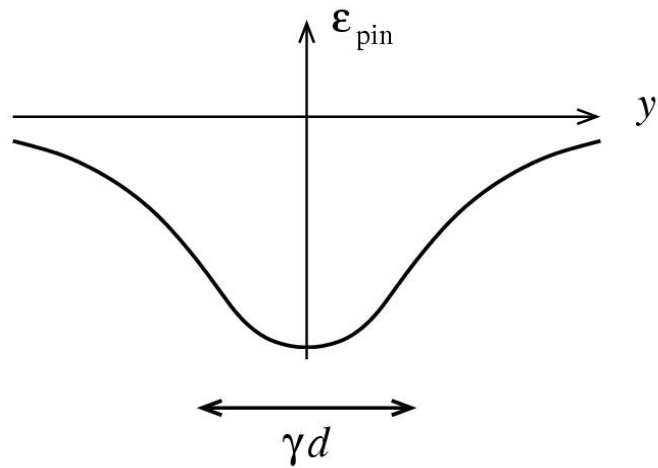

Figure 4.2.2  Schematic graph of the pinning energy profile of a PV stack displaced a distance $y$ from the centre of the JV stack.  At $y = 0$ this takes the value $\varepsilon_{pin} = E_\times / a_z$, while at distances greater than γd it takes the form $\varepsilon_{pin} \sim 1/y \ln y$  [see equations (4.2.4) and (4.2.17)].

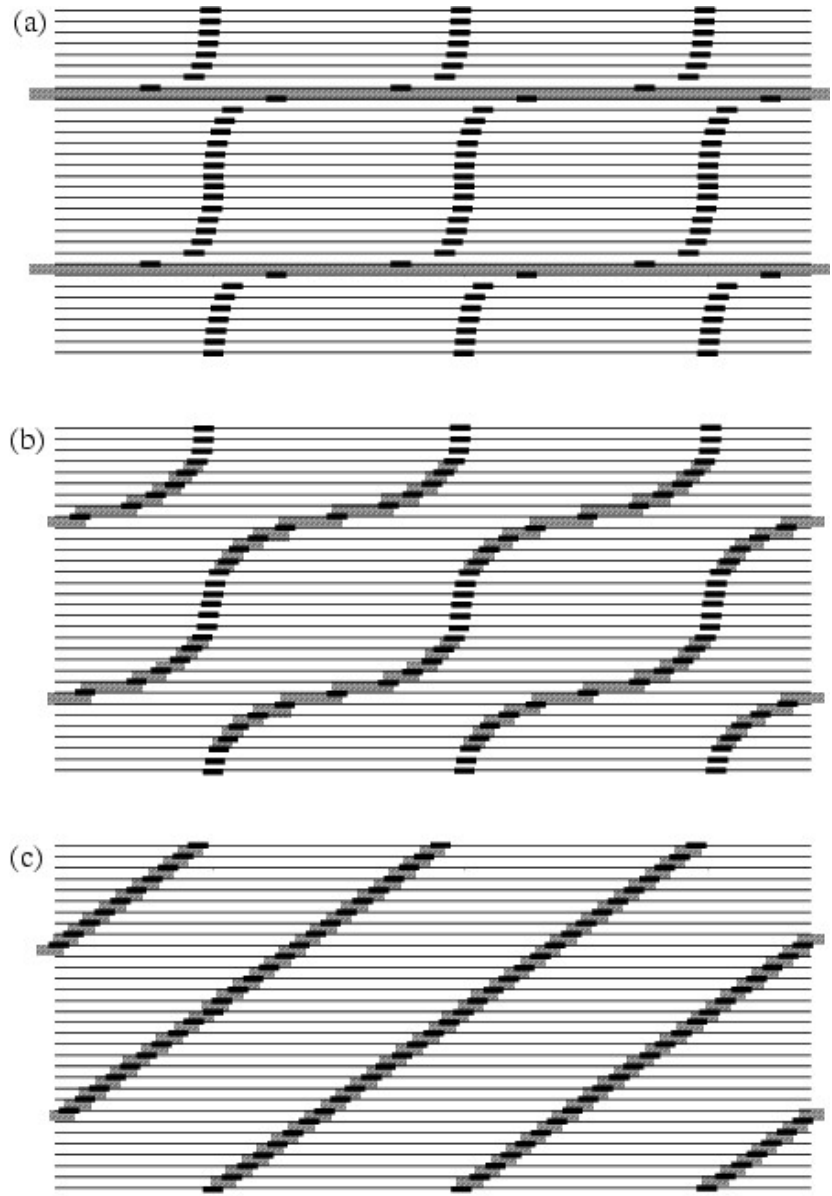

Figure 4.2.3  The three different structures that a chain may take, as elaborated by Koshelev:  (a) the crossing chain; (b) the modulated tilted chain; (c) the tilted chain.  All three types may be seen at very high anisotropy, for different values of the vortex spacing along the chain: On increasing density the crossing chain transforms smoothly to the modulated tilted chain, which then, at a higher density, has a continuous transition to the straight tilted chain.  For low enough anisotropy only the tilted chain is stable.  In an intermediate range of anisotropy the tilted chain first appears for an extremely dilute density.  There is then a sharp transition to a crossing chain of higher density and subsequently, on increasing the density, the modulated chain is seen and then finally the straight tilted chain reappears.

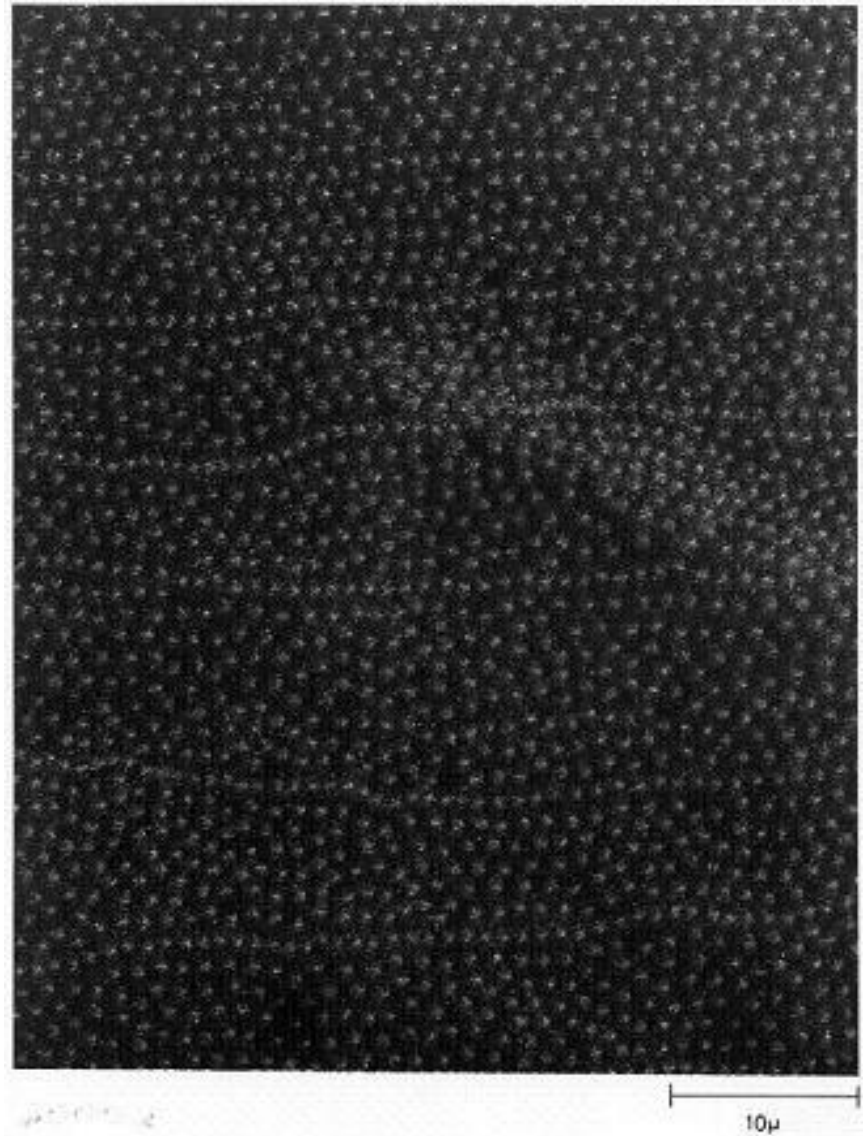

Fig. 4.3.1  Micrograph of a BSCCO single crystal which had been Bitter decorated after field-cooling to low temperature in an applied  field of 35 Oe tilted 70$^{o}$ away from the crystalline c-axis. The field of view is 60μm × 75μm.  [*Reproduced with permission from reference [32]. Copyright 1991 American Physical Society*]

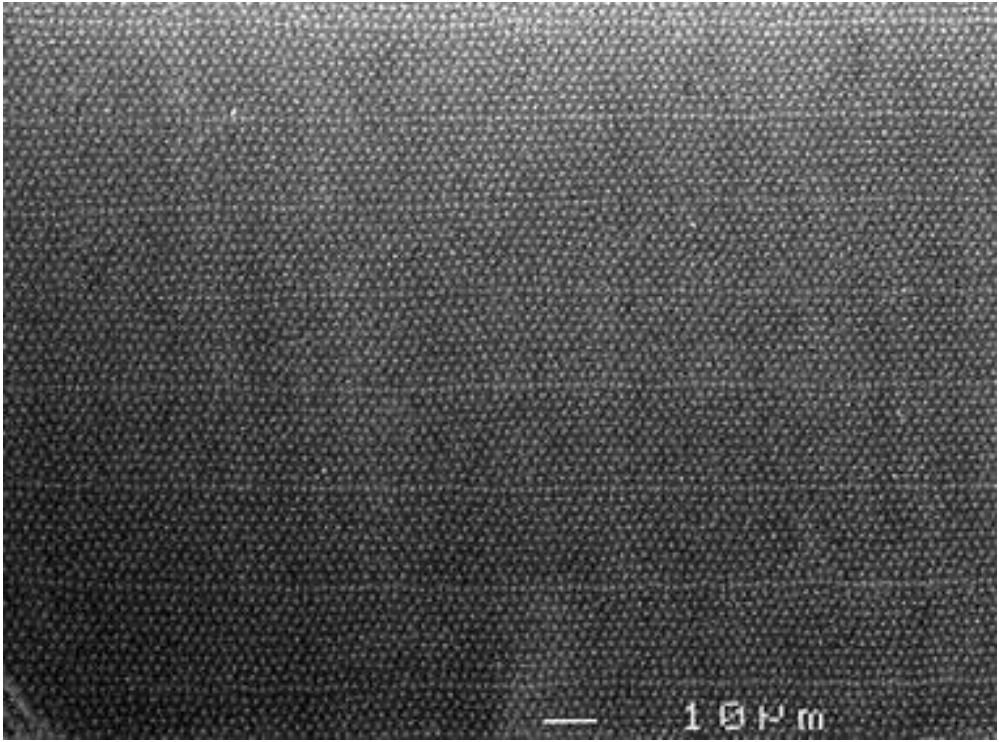

Fig. 4.3.2  Micrograph of a BSCCO single crystal which had been Bitter decorated are field-cooling to 4.2 K in an applied  field of 20 Oe tilted 70$^o$ away from the crystalline c-axis.  [*Image courtesy of Dr I.V.Grigorieva, University of Manchester, UK*]

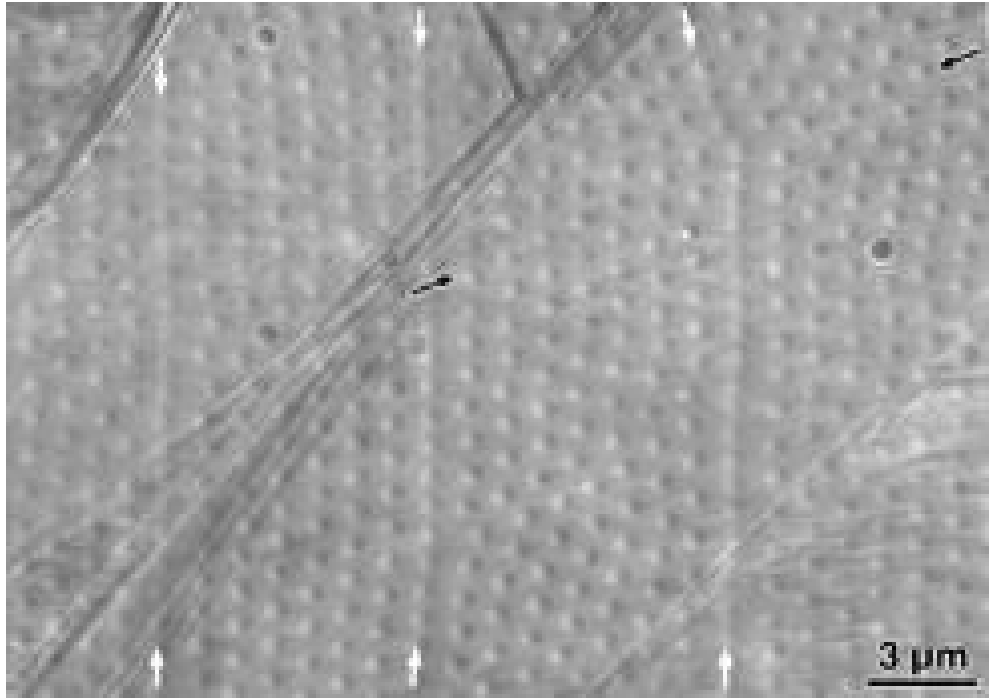

Fig. 4.3.3  Lorentz micrograph of the composite chains/lattice state in a very thin BSCCO crystal at 50K with a field of 58Oe applied at 80° away from the crystalline c-axis.  White arrows show the locations of vortex chains, black arrows indicate a surface step.  [*Reproduced with permission from reference [50]. Copyright 2001 AAAS*].

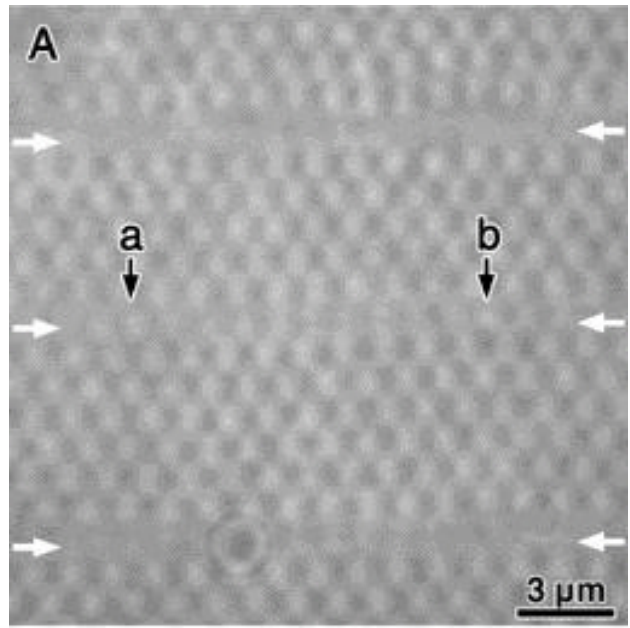

Fig. 4.3.4 Lorentz micrograph of the composite chains/lattice state in a very thin BSCCO crystal at 70K with a field of 69Oe applied at 80° away from the crystalline c-axis. White arrows show the location of vortex chains. The image illustrates the loss of vortex contrast in the chains except for those PV stacks labelled a and b in the central chain. [*Reproduced with permission from reference [50]. Copyright 2001 AAAS*].

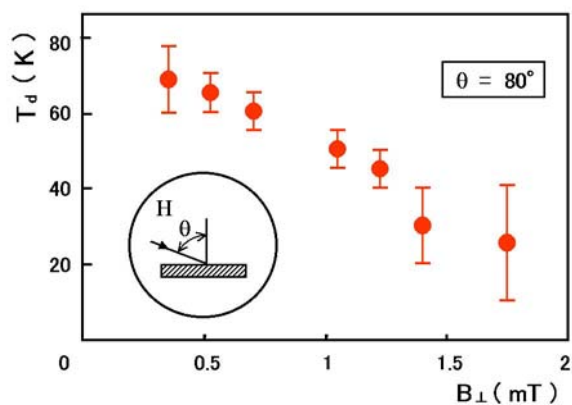

Fig. 4.3.5 Graph showing dependence of the disappearance temperature, $T_d$, on the perpendicular component of magnetic field. [*Reproduced with permission from reference [50]. Copyright 2001 AAAS*].

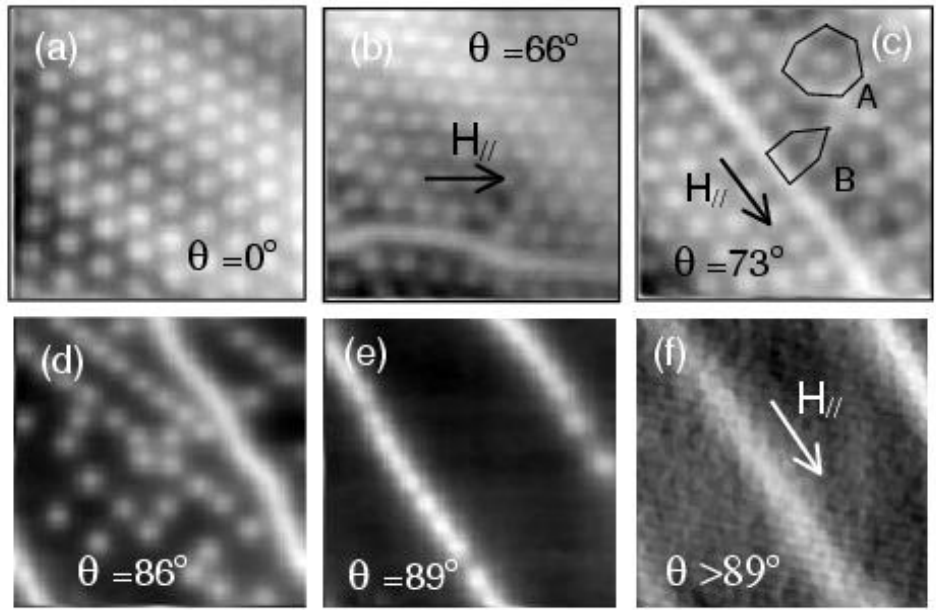

Fig. 4.4.1  SHPM images of: (a) The hexagonal Abrikosov lattice of pancake vortices obtained at $H_z$=12Oe ($H_\parallel$=0) at 81K.  (Grayscale spans 2.5G.).  The composite chains/lattice state at (b) $H_z$=14Oe and $H_\parallel$=32Oe at 81K (GS=1.7G) and (c) $H_z$=10Oe and $H_\parallel$=32Oe at 77K (GS=1.9G).   (d) The composite state below the "ordering" field at $H_z$=2Oe and $H_\parallel$=27Oe at 81K (GS=2.9G).  (e) The "isolated chain" state at $H_z$=0.5Oe and $H_\parallel$=35Oe at 81K (GS=2.5G).  (f)  The re-entrant uniformly tilted vortex state at $H_z$<0.2Oe and $H_\parallel$=38Oe at 83K (GS=0.4G).  [*Reproduced with permission from reference [47]. Copyright 2001 Nature Publishing Group*].

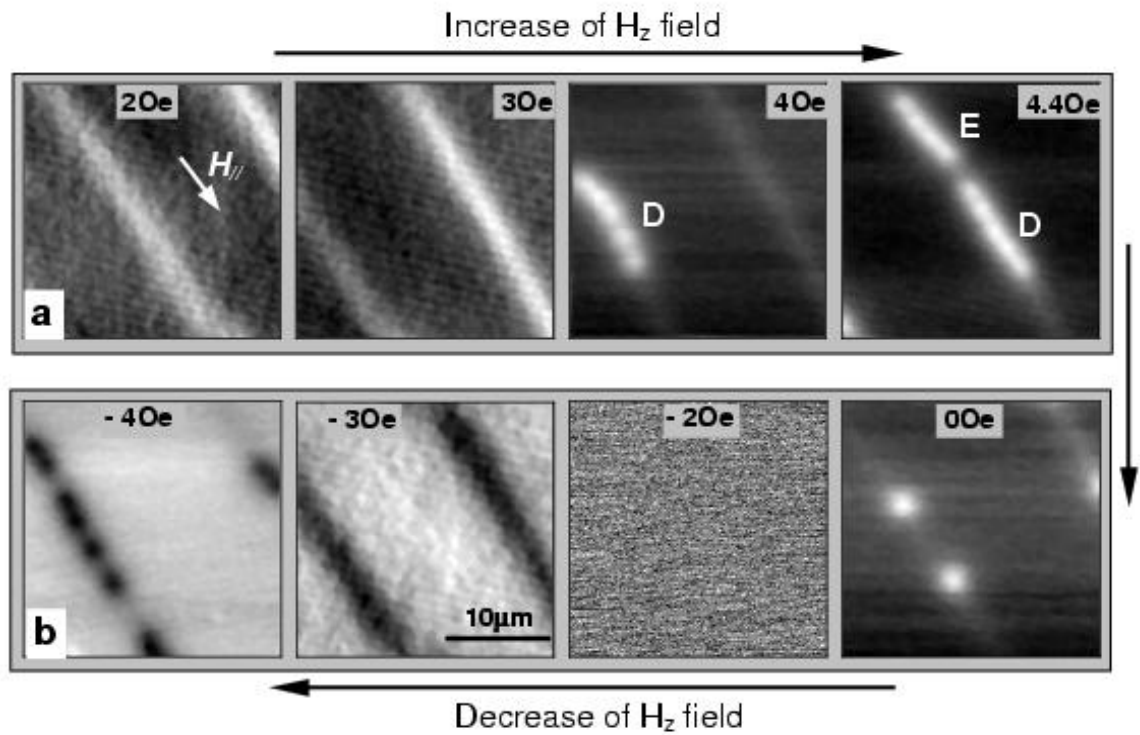

Figure 4.4.2  Images of vortices in a BSCCO single crystal for a fixed in-plane field $H_{\parallel}$=38Oe at T=83K. **a** ($H_z$ increasing) The tilted vortex state at $H_z$ =2Oe, (GS=0.35G) and $H_z$ =3Oe, (GS=0.4G) as well as this state coexisting with small segments of interacting PV stacks at $H_z$ =4Oe, (GS=2.3G) and H$_z$ =4.4Oe, (GS=2.3G).  **b**  ($H_z$ decreasing) The segments of PV stacks shrink as the field is reduced to $H_z$ =0.0Oe, (GS=1.7G) and the total expulsion of flux at $H_z$ =-2.0Oe, (GS=0.14G) as well as the tilted lattice state in the reverse direction at $H_z$ =-3Oe, (GS=0.3G) and its coexistence with interacting PV stacks at $H_z$ =-4Oe, (GS=2.1G).  [*Reproduced with permission from reference [47]. Copyright 2001 Nature Publishing Group*].

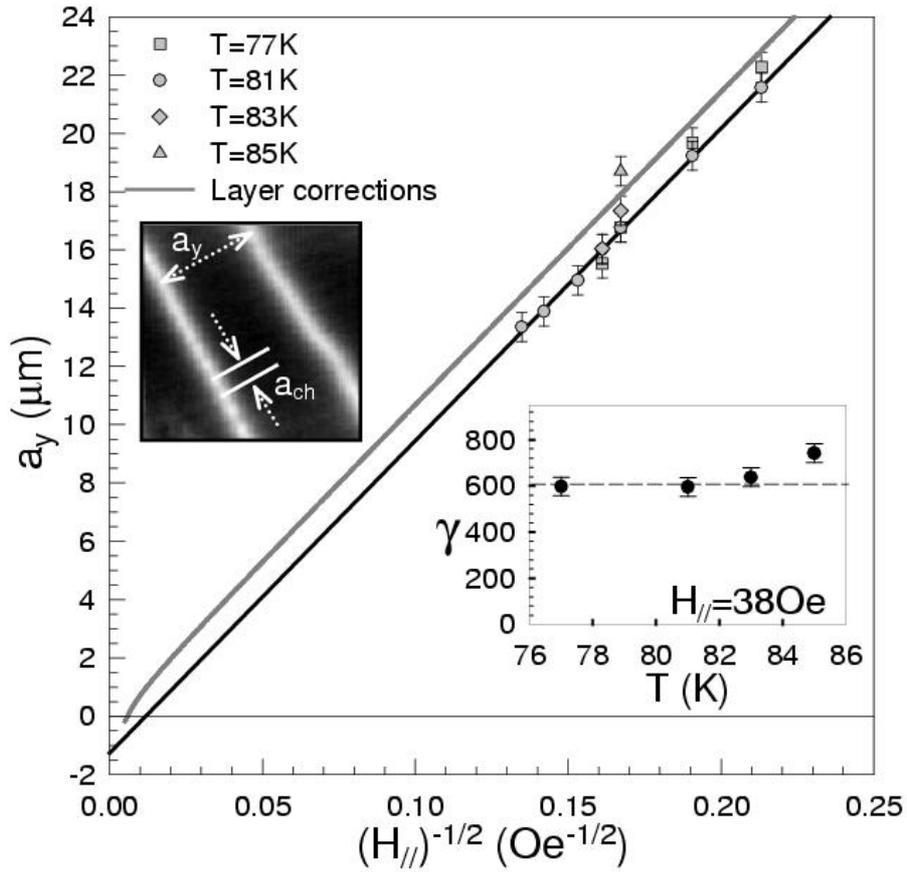

Fig. 4.4.3 Plot of the mean separation of 1D pancake vortex chains as a function of $(H_{\parallel})^{-1/2}$ at various temperatures. Left hand inset shows an image of pancake vortex chains in a BSCCO single crystal under a tilted magnetic field at 81K ($H_{\parallel}$=49.5Oe, $H_z$=0.7Oe, image size ~27μm×27μm). Right hand inset shows the temperature dependence of the anisotropy at $H_{\parallel}$=38Oe. [*Reproduced with permission from reference [54]. Copyright 2002 American Physical Society*]

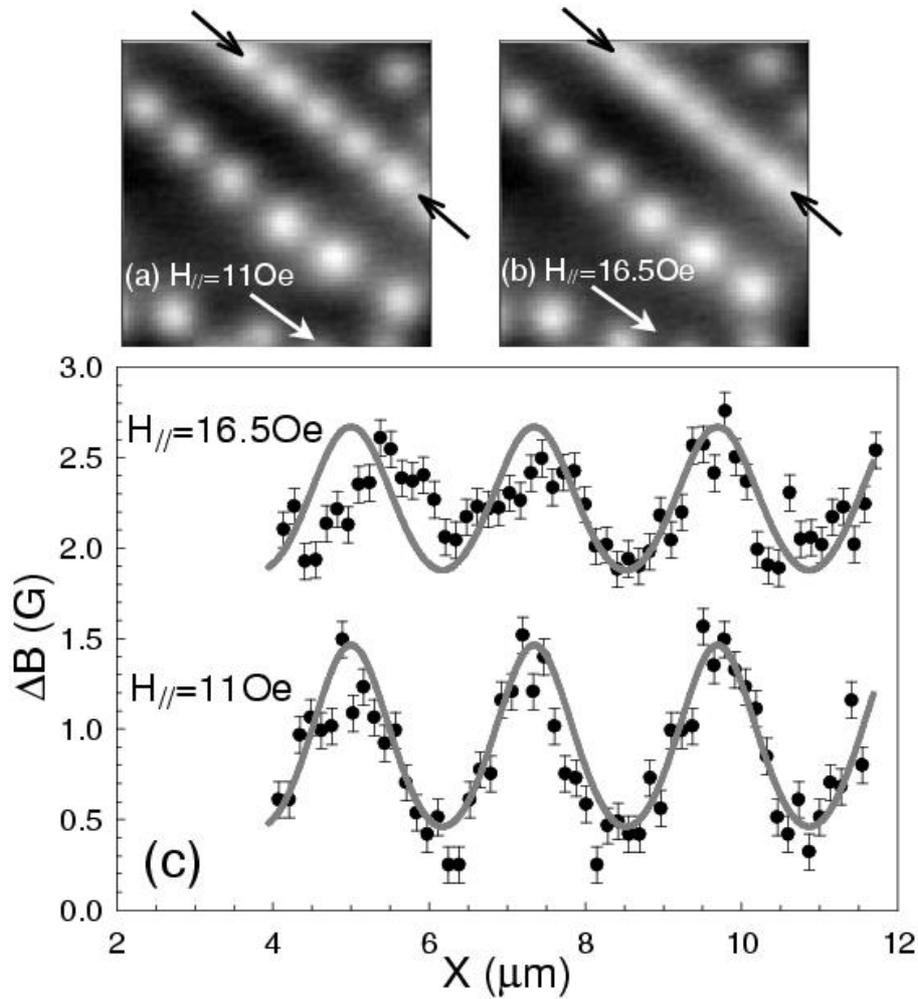

Fig. 4.4.4(a) Image of pinned pancake vortex stacks after field-cooling to 83K in $H_z$=1.8Oe ($H_\parallel$=0) followed by the application of a small in-plane field $H_\parallel$=11Oe along the crystalline a-axis (image size ~14μm×14μm). (b) Same region of the sample after the in-plane field was increased to $H_\parallel$=16.5Oe. (c) Linescans along the directions indicated in (a),(b). Smooth gray curves are fits to equation (3.3.2) for appropriate parameters (see text). [*Reproduced with permission from reference [54]. Copyright 2002 American Physical Society*]

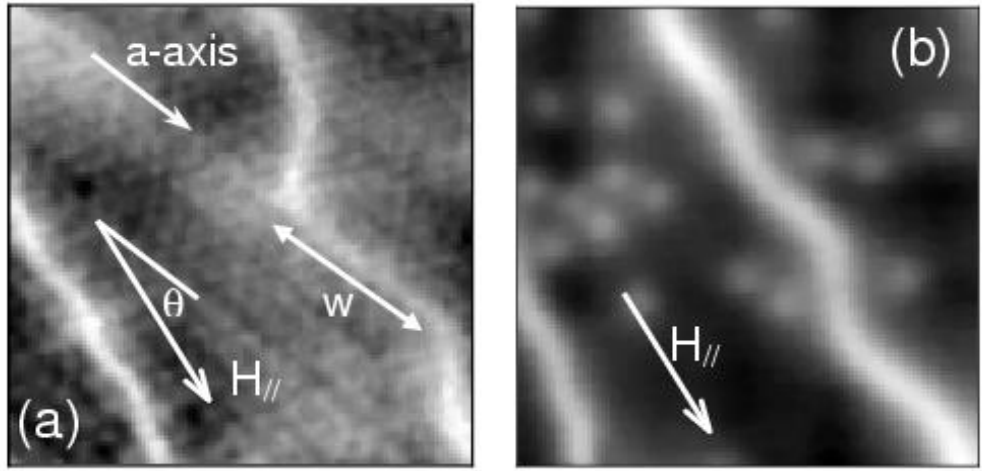

Fig. 4.4.5(a)  SHPM image of 'kinked' pancake vortex chains at 85K ($H_\parallel$=35.8Oe, $H_z$=8.8Oe, image size ~27.5μm×27.5μm).  (b)  Kinked chains at a lower *c*-axis magnetic field at 83K ($H_\parallel$=35.8Oe, $H_z$=4.4Oe, image size ~14μm×14μm).  [*Reproduced with permission from reference [54]. Copyright 2002 American Physical Society*]

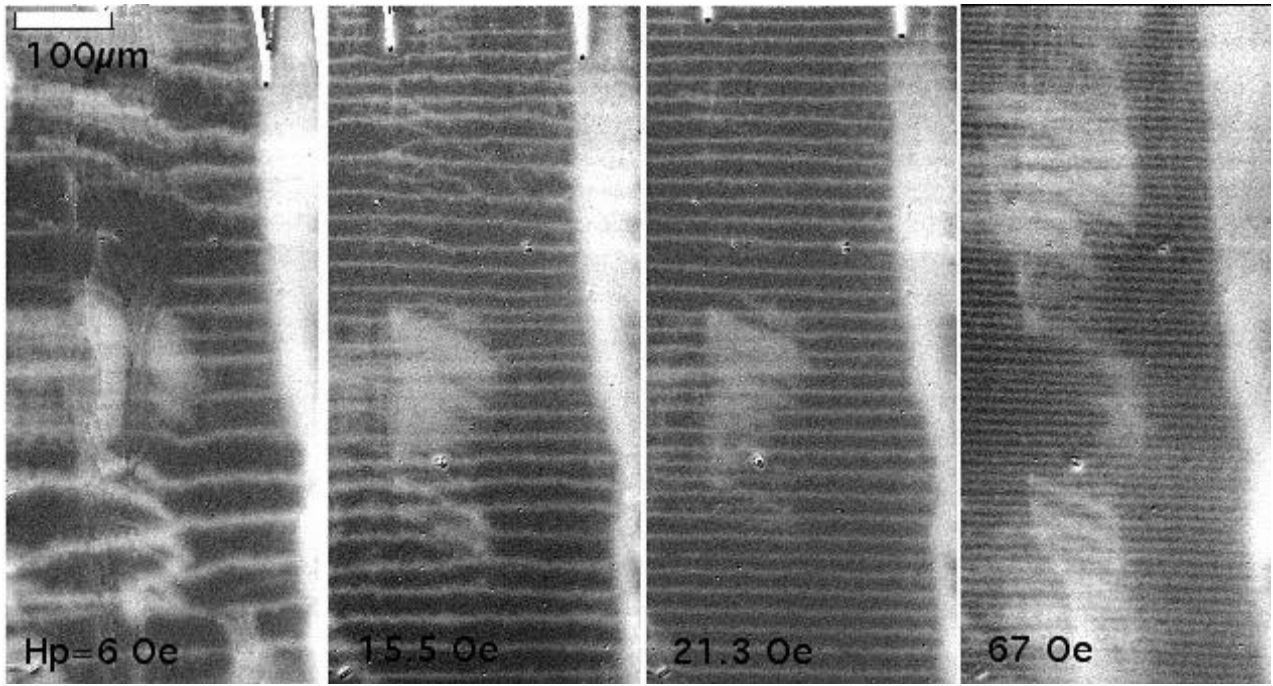

Fig. 4.4.6  Magneto-optical images of decorated JV stacks (white) in a BSCCO single crystal at the indicated in-plane fields.  $H_z$=2Oe, *T*=82K.  [*Reproduced with permission from reference [63]. Copyright 2003 Kluwer Academic Publishers*]

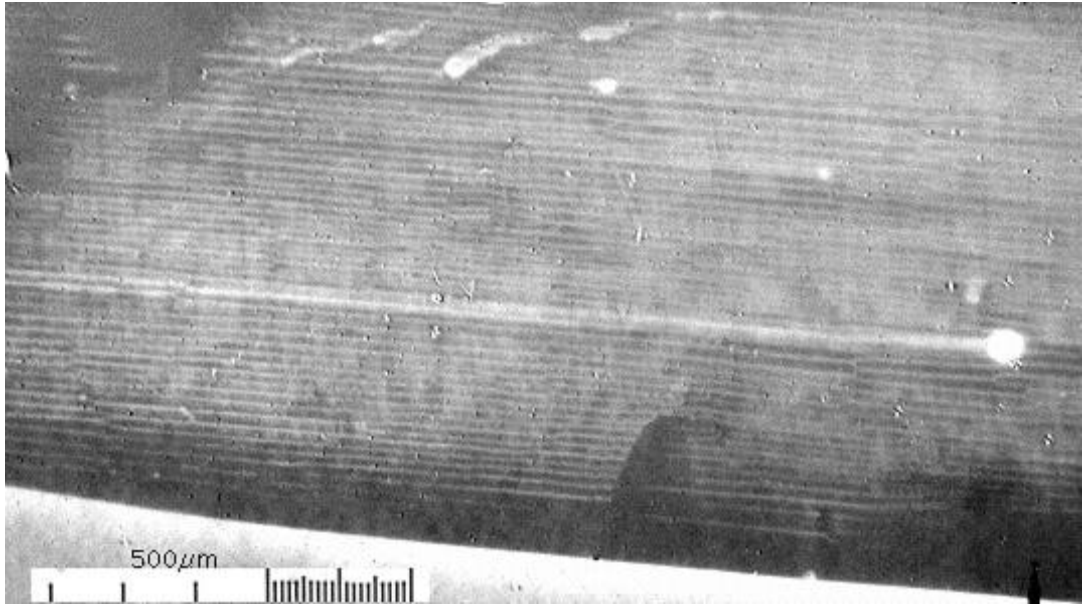

Fig. 4.4.7  Magneto-optical images of decorated JV stacks (white) over a large fraction of a BSCCO single crystal at 80K.  $H_\parallel$=36 Oe, and $H_z$= 2 Oe.  [*Reproduced with permission from reference [63]. Copyright 2003 Kluwer Academic Publishers*]

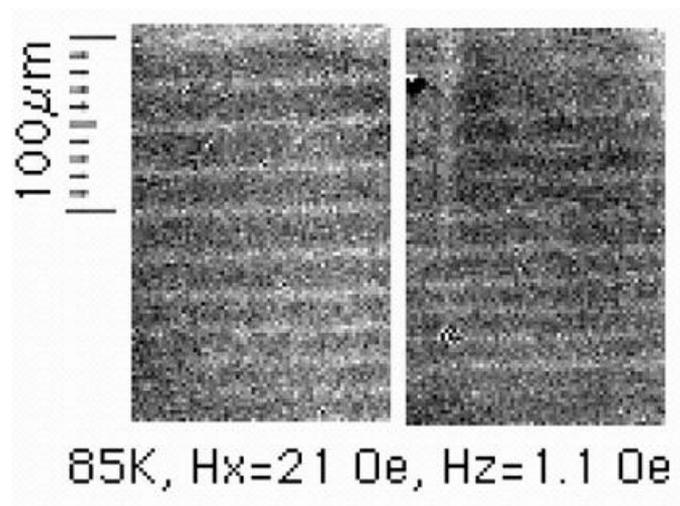

Fig. 4.4.8  Magneto-optical images of decorated JV stacks observed simultaneously in two distant regions of a BSCCO single crystal at 85K.  ($H_\parallel$=21 Oe, $H_z$=1.1 Oe).  [*Reproduced with permission from reference [63]. Copyright 2003 Kluwer Academic Publishers*]

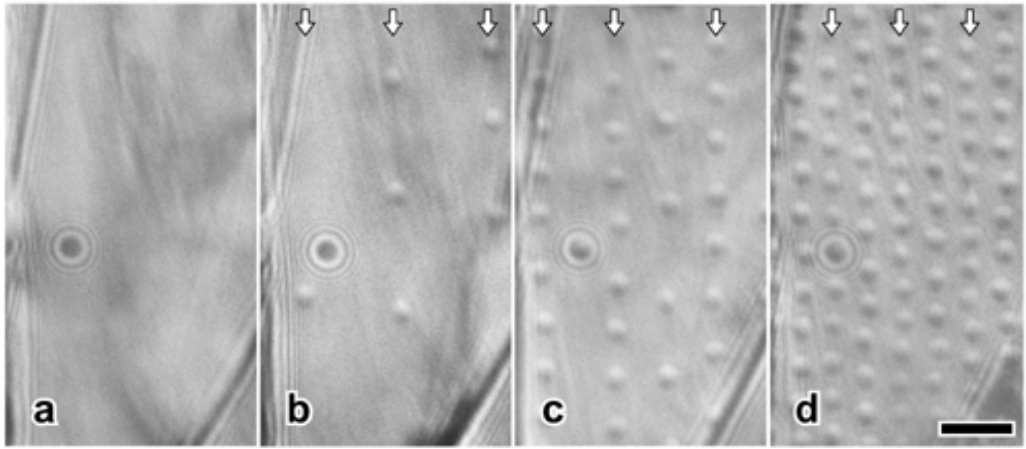

Fig, 4.4.9  Lorentz micrograph of PV stacks decorating underlying JV chains at $T$=50K and $H_\parallel$=5Oe at various perpendicular magnetic fields.  (a) $H_z$=0, (b) $H_z$=0.2Oe, (c) $H_z$=1Oe and (d) $H_z$=1.7Oe. White arrows show the inferred locations of some of the JV stacks.  [*Reproduced with permission from reference [20]. Copyright 2002 American Physical Society*]

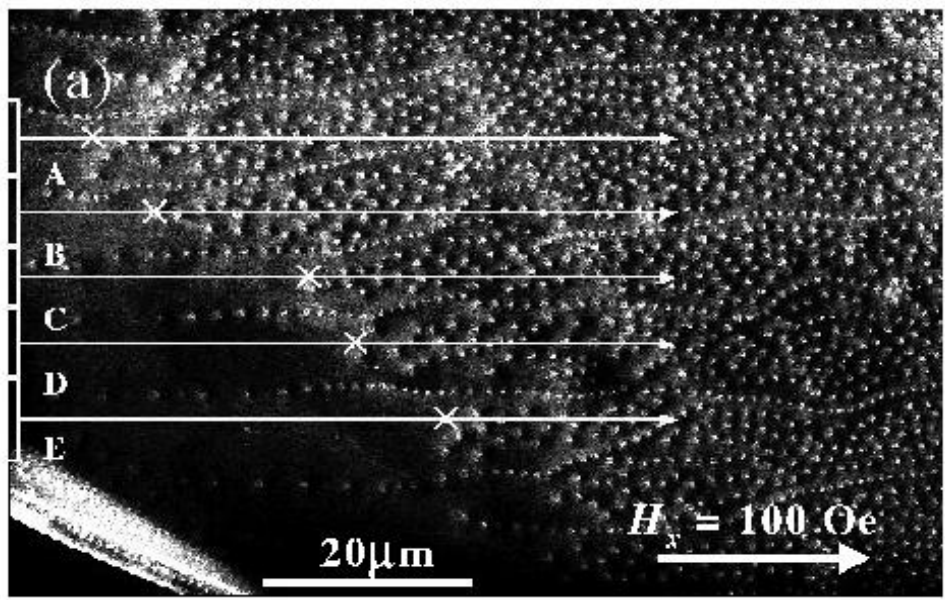

Fig. 4.4.10  Bitter decoration image at $H_\parallel$=100Oe and $H_z$=15Oe near the edge of a highly overdoped BSCCO crystal (see text for details of decoration).  [*Reproduced with permission from reference [64]. Copyright 2003 American Physical Society*]

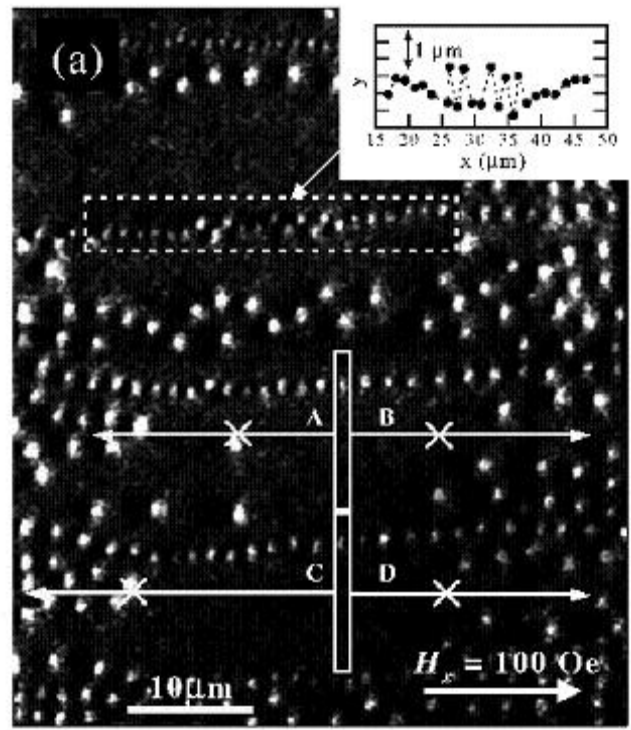

Fig. 4.4.11  Bitter decoration image of an optimally doped BSCCO crystal at $H_{\parallel}$=100Oe and $H_z$=10Oe.  Inset shows the estimated positions of PV stacks in the region indicated by the white dashed box.  [*Reproduced with permission from reference [64]. Copyright 2003 American Physical Society*]

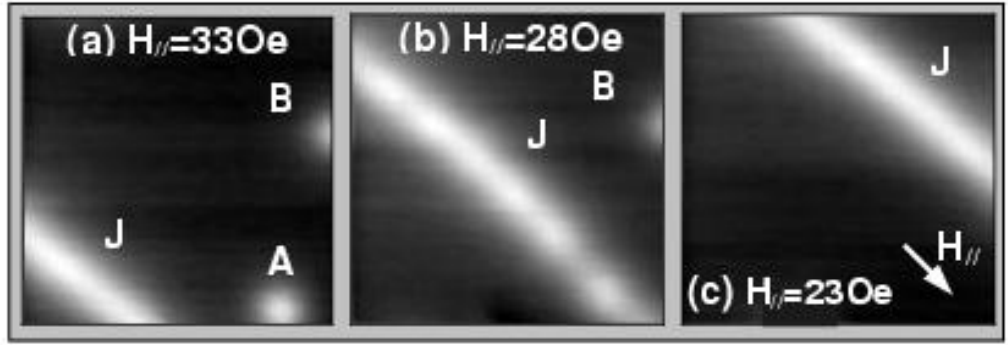

Fig. 4.5.1 Series of sequential SHPM images showing how a pancake vortex chain can be dragged by a JV stack at 83K as $H_\parallel$ is reduced at fixed $H_z$=0.5Oe. (a) $H_\parallel$ =33Oe (GS=2.9G), (b) $H_\parallel$ =28 Oe (GS=3.2G) and (c) $H_\parallel$ =23 Oe (GS=3.4G). A, B are PV stacks that have been picked up by the chain as it moves. [*Reproduced with permission from reference [47]. Copyright 2001 Nature Publishing Group*].

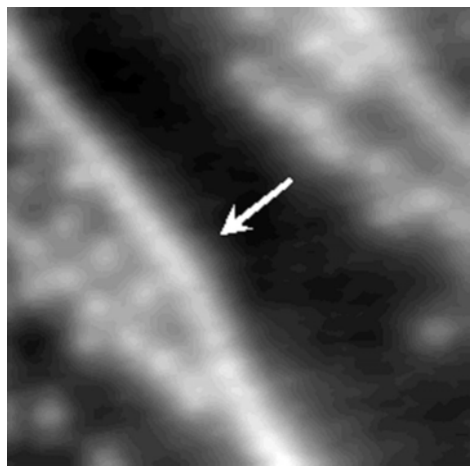

Fig. 4.5.2 SHPM image of a BSCCO single crystal which had been field-cooled in $H_z$=10Oe ($H_\parallel$=0), and then $H_z$ set to zero at $T$=77K. The in-plane field was then increased to $H_\parallel$=39Oe and the generated JV stacks have swept out the trapped PV flux in their direction of motion (white arrow). (Image size ~26µm×26µm).

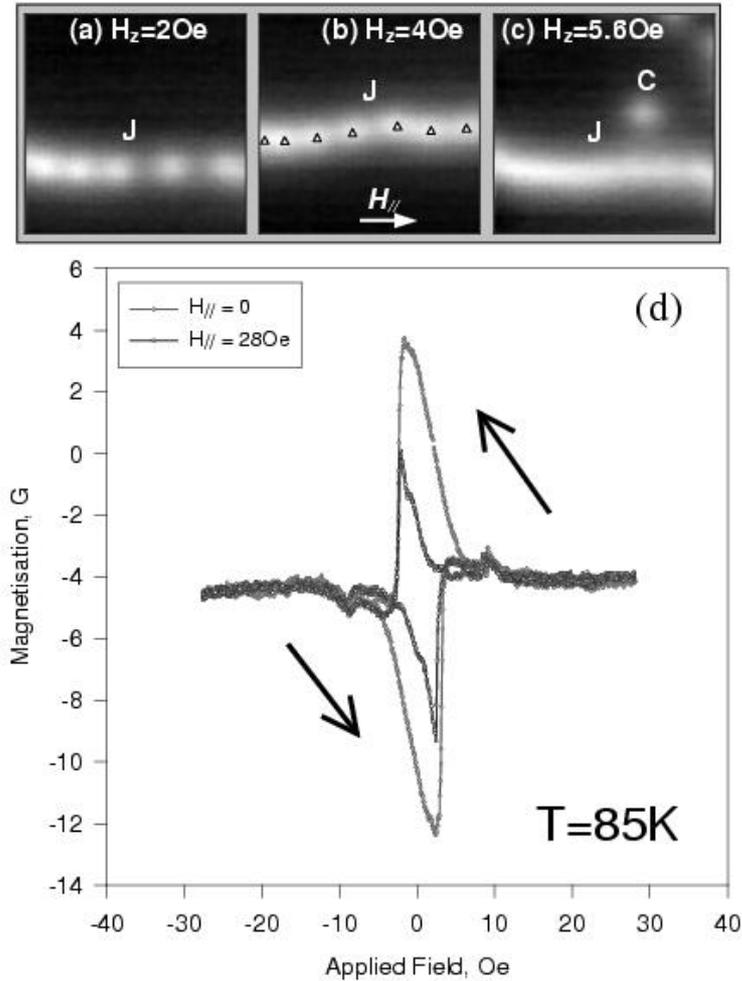

Fig. 4.5.3  SHPM images of penetration of PV stacks along 1D chains at 81K as $H_z$ is increased at fixed $H_\parallel$=35Oe.  (a) $H_z$ =2Oe, (GS=3.2G).  (b) $H_z$ =4Oe, (GS=3.7G).  (c) $H_z$ =5.6Oe, (GS=4.2G). (d)  Local magnetisation loop measured with the Hall probe ~1μm above the sample with ($H_\parallel$=28Oe) and without ($H_\parallel$=0) an in-plane magnetic field.  [*Reproduced with permission from reference [47]. Copyright 2001 Nature Publishing Group*].

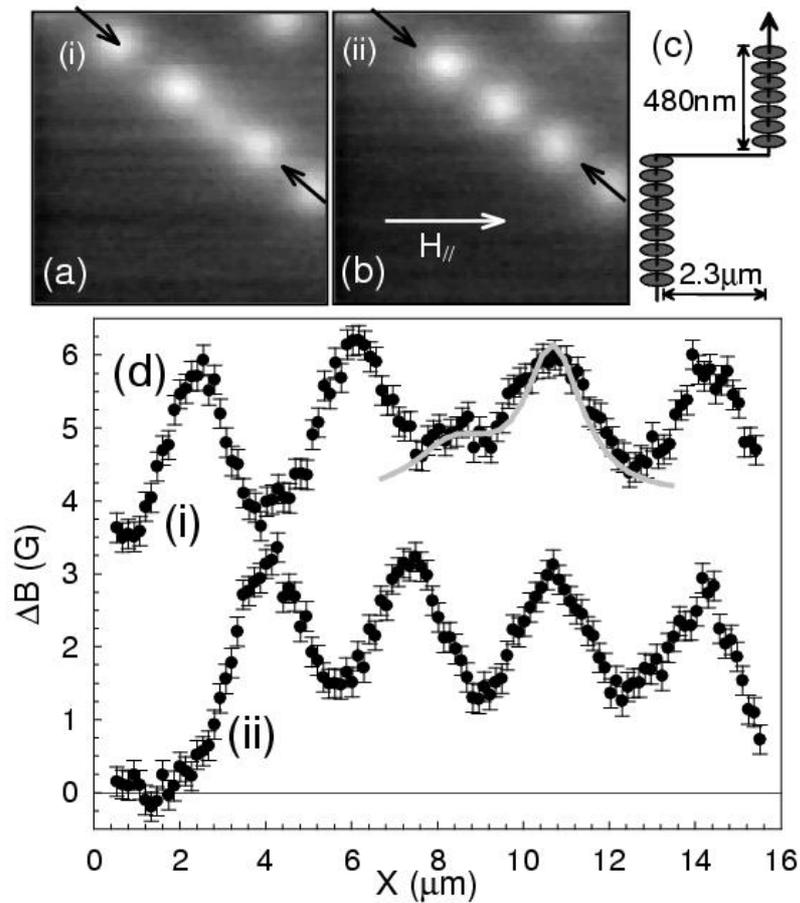

Fig. 4.5.4(a) SHPM image of a split PV stack ($H_z$ =0, T=85K) after the in-plane field was suddenly reduced from $H_\parallel$=36Oe to zero (image size ~14μm×14μm). (b) 'Healing' of split PV stack after the in-plane field was cycled back up to $H_\parallel$=11Oe. (c) Sketch of the split PV stack used to model linescans in (d). (d) Linescans along the indicated directions. The gray curve is a fit to the pancake vortex model (see text). [*Reproduced with permission from reference [54]. Copyright 2002 American Physical Society*]

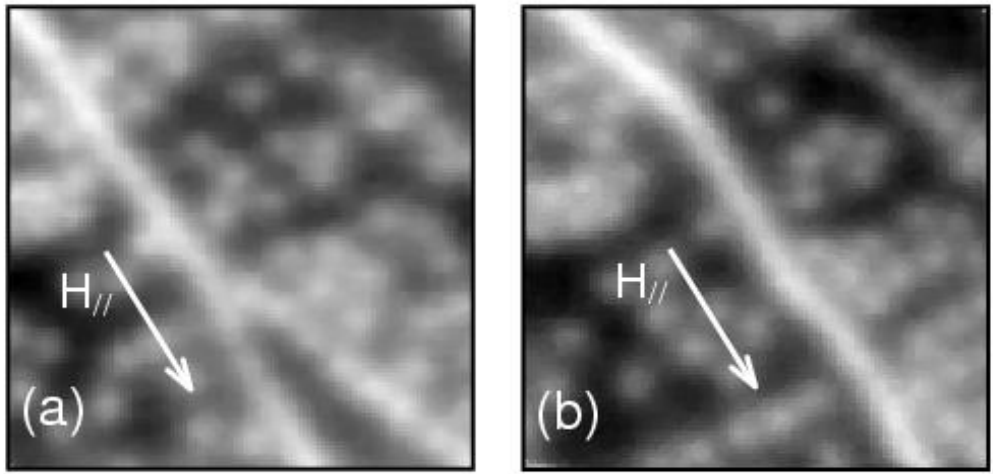

Fig. 4.5.5(a)  SHPM image of a decorated 'forked' JV stack after the in-plane field was abruptly increased to $H_\parallel$=33Oe at 77K ($H_z$=1Oe, image size ~26μm×26μm).  (b)  'Healing' of the fork observed in (a) after the in-plane field was reduced to $H_\parallel$=22Oe.  [*Reproduced with permission from reference [54]. Copyright 2002 American Physical Society*]